\documentclass[12pt]{article}
\usepackage[a4paper, margin = 1in]{geometry}
\date{}

\usepackage[utf8]{inputenc}
\usepackage{amssymb}
\usepackage{amsmath}
\usepackage{booktabs} 
\usepackage{mathtools}
\usepackage{bm}
\usepackage{graphicx}
\usepackage[table]{xcolor} 
\usepackage{natbib}
\setcitestyle{aysep={}}   

\usepackage{array}
\usepackage{arydshln}

\usepackage{xcolor, colortbl}
\newcommand{\W}{\boldsymbol{\Omega}}
\newcommand{\X}{\boldsymbol{X}}
\newcommand{\Y}{\boldsymbol{Y}}
\newcommand{\BR}{\boldsymbol{R}}

\newcommand{\bw}{\boldsymbol{\omega}}
\newcommand{\bx}{\boldsymbol{x}}
\newcommand{\by}{\boldsymbol{y}}
\newcommand{\br}{\boldsymbol{r}}

\newcommand{\bs}{\boldsymbol{s}}

\newcommand{\thetab}{\boldsymbol{\theta}}
\newcommand{\etab}{\boldsymbol{\eta}}
\newcommand{\phib}{\boldsymbol{\phi}}
\newcommand{\nub}{\boldsymbol{\nu}}
\newcommand{\gammab}{\boldsymbol{\gamma}}
\newcommand{\xib}{\boldsymbol{\xi}}
\newcommand{\E}{\mathbb{E}}

\newcommand{\red}{\textcolor{black}}

\title{Accommodating false positives within acoustic spatial capture-recapture, with variable source levels, noisy bearings and an inhomogeneous spatial density}

\author{Felix T Petersma$^{1}$\footnote{ftp@st-andrews.ac.uk}, 
Len Thomas$^1$, 
Aaron M Thode$^2$, 
Danielle Harris$^1$, \\ 
Tiago A Marques$^{1,3}$, 
Gisela V Cheoo$^3$, and
Katherine H Kim$^4$ \\ \\
$^{1}$Centre for Research into Environmental and Ecological Modelling, \\ University of St Andrews, St Andrews KY16 9LZ, U.K.\\
$^{2}$Scripps Institution of Oceanography, \\University of California San Diego,
San Diego, U.S.A. \\
$^{3}$Centro de Estat\'{i}stica e Aplica\c{c}\~{o}es, Departamento de Biologia Animal, \\ Faculdade de Ci\^{e}ncias da Universidade de Lisboa, Portugal \\ 
$^{4}$Greeneridge Sciences, Inc., 5142 Hollister Avenue, \\Suite 283, Santa Barbara, California 93111, U.S.A.}

\begin{document}

\maketitle

\begin{abstract}
Passive acoustic monitoring is a promising method for surveying wildlife populations that are easier to detect acoustically than visually.
When animal vocalisations can be uniquely identified on an array of sensors, the potential exists to estimate population density through acoustic spatial capture-recapture (ASCR).
However, sound classification is imperfect, and in some situations a high proportion of sounds detected on just a single sensor (`singletons') are not from the target species.  We present a case study of bowhead whale calls (\textit{Baleana mysticetus}) collected in the Beaufort Sea in 2010 containing such false positives.
We propose a novel extension of ASCR that is robust to false positives by truncating singletons and conditioning on calls being detected by at least two sensors.
We allow for individual-level detection heterogeneity through modelling a variable sound source level, model inhomogeneous call spatial density, and include bearings with varying measurement error. 
We show via simulation that the method produces near-unbiased estimates when correctly specified. 
Ignoring source level variation resulted in a strong negative bias, while ignoring inhomogeneous density resulted in severe positive bias. 
The case study analysis indicated a band of higher call density approximately 30km from shore; 59.8\% of singletons were estimated to have been false positives. 

\red{Supplementary materials accompanying this paper appear online.}
\end{abstract}

%



\maketitle

\section{Introduction}
\label{sec:intro}

In recent decades, passive acoustic monitoring (PAM) has increasingly been used to study both terrestrial \citep{Sugai2019Terrestrial} and marine animals, particularly cetaceans \citep{Zimmer2011PassiveAcoustic}. 
Compared with more traditional visual survey methods, acoustic monitoring works day and night, is robust to variation in environmental conditions such as weather, and in some habitats has the potential to detect animals at greater distances, hence increasing the area surveyed \citep{Marques2011Estimating}. 
It has enabled studies of rare and elusive species such as the vaquita \citep{Thomas2017LastCall} and several species of beaked whale \citep{Hildebrand2015Passive, Yack2013Passive} which, despite being visually cryptic, produce frequent sounds that can be detected by PAM systems.

One important application of PAM is to estimate population density or abundance \citep{Marques2013}.  In the situation where multiple acoustic sensors are deployed simultaneously with a spatial separation such that some vocalisations can be detected on multiple sensors then an appropriate statistical framework for estimating density is spatial capture-recapture (SCR; also known as spatially explicit capture-recapture)  \citep[e.g.,][]{Borchers2015Unifying, Stevenson2015}. 
SCR is an extension of long-established capture-recapture (otherwise known as mark-recapture or mark-resight) methods where data on the detection (`capture') of individually-identified animals is supplemented by data on the spatial location of the survey effort and the detections \citep{Borchers2008Spatially, Royle2009Hierarchical, Kidney2016Efficient}. 
Recording not just whether but also where each animal was detected increases the accuracy of abundance and density estimates and potentially allows estimation of a spatially inhomogeneous animal density surface. 

Acoustic spatial capture-recapture (ASCR) is a special case of SCR where the detections are of individual animal vocalisations or `cues'.  
Standard SCR relies on animals moving between detection locations and hence typically requires multiple capture occasions, while in ASCR the sound travels almost instantaneously from its source and hence can be detected on multiple sensors in a single occasion.  
Estimation is of cue spatial density; to convert to animal density an estimate of average cue production rate is required \citep{Marques2013, Stevenson2021Spatial}.  As well as the location of detection, additional information is often available about the location of the vocalisation, e.g., the bearing, received sound level or time of arrival on multiple sensors. \cite{Borchers2015Unifying} showed that using this additional information further improved estimation accuracy.  

PAM data can also present challenges that require accounting for to avoid bias in ASCR analysis. 
First, sound classification is imperfect, leading to false positive detections of sounds not originating from the target species. 
Second, vocalisation source level can vary considerably, causing heterogeneity in detectability. 
Third, there can be considerable measurement error in the additional information, particularly the bearings.  
Forth (in common with other SCR studies), spatial density of source locations can vary substantially.

The methods presented here are motivated by a case study that demonstrates all four of the above issues: estimation of call density of bowhead whales (\textit{Baleana mysticetus}) migrating through the Beaufort Sea. 
Multiple arrays of acoustic sensors were deployed in the Beaufort Sea during the migration season and recorded millions of bowhead whale calls.
Automated detection and classification methods were therefore used to process the data, yielding call detections, received sound levels and bearings, and linking calls across detectors.  
However, a high proportion of the detections made on only one sensor (`singletons') were thought to be false positives. 
\red{Naively including these singletons in the analysis would lead to a positive bias in the estimated abundance of unknown but likely substantial magnitude.}
To avoid this, we excluded all singletons from the analysis and conditioned the ASCR likelihood to only include calls detected on multiple sensors (`multiply-detected calls').  
\red{Truncating the data in this manner, rather than attempting to model the proportion of false positives, is a good strategy when data are plentiful (see Discussion).}
\cite{Conn2011Accounting} proposed a similar procedure in a mark-resight study as a way to differentiate between resident and transient bottlenose dolphin populations, assuming that transients were never detected more than once.  
To our knowledge, this is the first time this approach has been used in SCR.

Our case study features some additional complications that may commonly appear in real-world data but are not typically all dealt with. 
Call source level was thought to vary substantially and so we include source level as a random effect in our model.  
Exploratory analysis showed that, while most estimated bearings were accurate, some appeared to be very inaccurate.  We therefore developed a two-part discrete mixture model for bearing error, extending the bearing error methods of \cite{Borchers2015Unifying}.  
Finally, we allowed for an inhomogeneous density model to accommodate the spatial preference of the migrating whales (and thus their calls).

In Section 2 we describe the case study in more detail.  
We then present the extended ASCR model in Section 3.  
In Section 4 we evaluate the model performance via simulation, and show how ignoring some of the real-world issues can result in substantial bias.  
Section 5 gives results from a proof-of-concept application of the model to a single day of data.   
Finally, in Section 6, we discuss results, limitations and alternatives.

\section{Case Study}
\label{sec:case_study}

Every year from August to October, the Bering–Chukchi–Beaufort (BCB) population of bowhead whales (\textit{Baleana mysticetus}) migrates westwards through the Beaufort Sea to their wintering areas in the Bering Sea \citep{Blackwell2007BowheadWhale}. 
They travel mainly over the continental shelf in waters less than 25 meters deep, approximately 10--75 km offshore \citep{Greene2004Directional}. 
During this migration, bowhead whales are known to produce a wide variety of calls \citep{Ljungblad1982Underwater}. 
The purpose of these calls remains largely unknown, although they may be used for long-range communication \citep{Thode2020Roaring} or to navigate through ice \citep{George1989Observations}. 

Between 2007 and 2014, up to 35 Directional Autonomous Seafloor Acoustic Recorders \cite[DASARs;][]{Greene2004Directional} were deployed at several sites off the north coast of Alaska to monitor the calling behaviours of the migrating whales during seismic surveys. 
An automated detection and classification procedure was developed to handle the more than one million detected calls over the period 2007--2014 \citep{Thode2012aAutomated, Thode2020Roaring}. 
This procedure could identify discrete sounds as bowhead whale calls, and subsequently link them with detections from other DASARs within the array if they were the same call \citep{Thode2012aAutomated}. 

\begin{figure} 
\centering
    \includegraphics[width=\columnwidth]{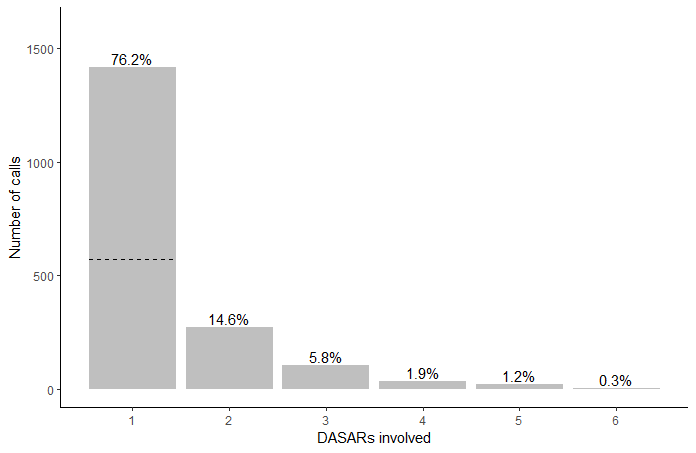}
    \caption{Counts of detected calls by number of DASARs (sensors) involved per call at site 5 on 31 August 2010. \red{Detections were included if the received level was at least 96 dB.} The high proportion of singletons (detections on a single DASAR; 1417) raised concerns about the validity of those calls.  The dashed line shows the number of singletons (570) that were estimated to be valid using the methods developed in this paper.}
    \label{fig:percentage_dasars_involved}
\end{figure}

Even though it was not the original purpose of the monitoring, the detection and linking of calls created detection histories for every detected call, making these data suitable for ASCR. 
ASCR theory assumes capture histories without errors, i.e., calls can be missed but not incorrectly positively identified or matched.
Several data cleaning procedures were required to meet this assumption as far as possible.
Sometimes, calls would be wrongly identified as other discrete sounds. 
\red{Distant airgun signals were occasionally misidentified as bowhead whale calls; bearded seals (\textit{Erignathus barbatus}) and walruses (\textit{Odobensu rosmarus divergens}) could appear similar as well, but these were generally rare and much quieter \citep{Thode2012aAutomated}.}
Moreover, if bowhead whale calls overlapped in time, they could be incorrectly matched as being the same call. 
\cite{Cheoo2019Estimation} showed that the number of detection histories that involved just one sensor (`singletons') in the automated data was not in line with expectations based on sound propagation theory (see Figure \ref{fig:percentage_dasars_involved}). 
It was therefore hypothesised that these singletons contain a high-degree of incorrectly classified calls (`false positives')\red{, mostly consisting of random fluctuations in the noise field or reverberations of whale calls, which were unlikely to be associated among multiple detectors.}
The solution we present here is to exclude singleton detection histories all together and modify the ASCR likelihood to be conditional on the capture histories involving at least two sensors.
\red{
To ensure that the multiply-detected calls would contain no false positives, we cleaned the remaining data using a procedure described in Appendix A in Supplementary Materials.}


In this study, we focus on the data from one specific day, 31 August 2010, from site 5, the most eastward site, as this site was one of the more intensely studied subsets of the data. 
Moreover, calls accumulated somewhat evenly across the day, resulting in a low expected rate of overlapping calls. 
For every call the following data were recorded: a detection history as well as bearings, received sound levels, and noise levels for every involved DASAR. 
The source level and origin of the call are unobserved, and hence treated as latent variables.
Throughout this manuscript, all sound measurements are denoted on the decibel scale (RMS; dB re 1 $\mu$Pa). 
More details on the background, availability, and pre-processing of the data are presented in Appendix A in Supplementary Materials. 

\section{Model}
\label{sec:model}
In this section, we first introduce the full likelihood. 
Following that, we derive each element of the likelihood individually.
Consider an acoustic survey of an array of $K$ sensors operating within a survey region $\mathcal{A} \subset \mathbb{R}^2$ over a period of time $T$.  
A total of $n$ unique animal calls are detected by at least two sensors.
For any multiply-detected call $i \in {1,...,n}$, let $\omega_{ij}$ be 1 if the call was detected at sensor $j \in {1,...,K}$, and $0$ otherwise. 
We define the matrix containing all detection histories as $\W = (\bw_1,...,\bw_n)^\top$, where the detection history for call $i$ is denoted $\bw_i = (\omega_{i1},...,\omega_{iK})^\top$ and $\top$ denotes the transpose.
For every call $i$ that was detected at sensor $j$, we also observe the bearing $y_{ij}$, measured in degrees clockwise relative to true north, and the received sound level $r_{ij}$.
These data are contained in $\Y = (\by_1,...,\by_n)^\top$ and $\BR = (\br_1,...,\br_n)^\top$, respectively. 
Latent variables are the spatial origins of calls, denoted by $\X = (\bx_1,...,\bx_n)^\top$ where $\bx_i$ is a location in the Cartesian plane, and the source levels, denoted by $\bs = (s_1,...,s_n)^\top$. 
The support for $s$ is denoted $\mathcal{S}$. 
Throughout this manuscript, we do not distinguish explicitly in notation between random variables and specific observations/realisations -- this should be clear from the context.

The parameter vectors used in the model are (following previous literature as closely as possible):  $\phib$ for parameters associated with the spatial density of emitted calls, $\nub$ for those associated with the source levels of calls, $\etab$ for those associated with source sound propagation and received levels, $\thetab$ for those associated with detectability of calls, and $\gammab$ for those associated with measurement error of the bearings.  
For notational convenience we define the joint parameter vector $\xib = (\phib, \nub, \etab, \thetab, \gammab)$. A list of model assumptions is presented in Section \ref{sec:assumptions}. 

\subsection{Likelihood Specification}
\label{sec:likelihood_specification}

The likelihood is formed from the joint distribution of all observed and latent variables introduced above. 
We denote this likelihood $\mathcal{L}_f(\xib)$ where the subscript $f$ stands for `full' to distinguish it from the conditional likelihood $\mathcal{L}_c(\xib)$ which is conditional on the observed number of detections.
Denoting both probability mass and density functions as $f(.)$, we define the full likelihood as
\begin{equation}
    \begin{aligned}
        \mathcal{L}_f(\xib) &= f(n, \W, \Y, \BR, \X, \bs; \xib) \\
        &= f(n ; \phib, \thetab, \nub) \times f(\W, \Y, \BR, \X, \bs| n; \xib) \\
&= f(n ; \phib, \thetab, \nub) \times \mathcal{L}_c(\xib).
    \end{aligned} \label{eqn:fulllikelihood}
\end{equation}
As we do not observe the call locations and source levels, we marginalise over the unobserved $\X$ and $\bs$ to obtain
\begin{equation}
    \begin{aligned}
        \mathcal{L}_c(\xib) 
        =& 
        \int_\mathcal{S}\int_{\mathcal{A}} 
        f(\W, \Y, \BR, \X, \bs| n; \xib) d\X d\bs. 
    \end{aligned}
    \label{eqn:conditionallikelihood}
\end{equation}
The double integral in Equation \eqref{eqn:conditionallikelihood} is of dimension \red{$3n$}, making this likelihood intractable. 
We follow \cite{Stevenson2015} in assuming calls to be independent of each other in all respects, allowing us to specify Equation \eqref{eqn:conditionallikelihood} as the product of \red{$n$ 3-dimensional} integrals:
\begin{equation}
    \begin{aligned}
        \mathcal{L}_c(\xib) \equiv &  \prod_{i \in \{1:n\}} \int_\mathcal{S}\int_{\mathcal{A}} f(\bw_i, \by_i, \br_i, \bx_i, s_i| \omega_{i}^* \geq 2; \xib) d\bx_i ds_i \\
        =& \prod_{i \in \{1:n\}} \int_\mathcal{S}\int_{\mathcal{A}} 
        f(\bw_i| \bx_i, \bs_i,  \omega_{i}^* \geq 2; \thetab) \\
        &\quad\quad\times f(\by_i| \bw_i, \bx_i; \gammab) \times 
        f(\br_i| \bw_i, \bx_i, s_i; \etab) \\
        & \quad\quad \times 
        f(\bx_i | s_i,  \omega_{i}^* \geq 2;  \phib, \thetab) \\ 
        & \quad\quad\times f(s_i|  \omega_{i}^* \geq 2 ;  \phib, \thetab, \nub) d\bx_{i}ds_i 
    \end{aligned}
    \label{eqn:condliktractable}
\end{equation}
where we assume independence between $\by_i$ and $\br_i$ given $\bx_i$, and $\omega_{i}^{*} := \sum_{j \in 1:k}w_{ij}$ denotes the total number of sensors involved in the detection of call $i$.
\red{The separation of the joint distribution inside the integrals in \eqref{eqn:condliktractable} follows from repeatedly applying Bayes' formula.}
Note that conditioning the joint distribution on $n$ in \eqref{eqn:conditionallikelihood} is equivalent to conditioning every marginal distribution on involving at least two sensors in \eqref{eqn:condliktractable}\red{; for the second and third element this conditioning is implicit in conditioning on detection history $\bw_i$.}
If we were to assume a constant spatial density of calls, it would be sufficient to simply maximise $\mathcal{L}_c$, as the parameter estimates from the conditional MLE are identical to those obtained by maximising the full likelihood -- a Horvitz Thompson-like estimator could then be used to derive the mean density \citep{Borchers2008Spatially}. 
In our case study, however, the  spatial density of calls is known to be inhomogeneous, so the full likelihood is required. 

In the following sections, we specify in further detail $f(n ; \phib, \thetab, \nub)$ and the five components in Equation \eqref{eqn:condliktractable}. 
For readability we will hereforth omit the indexing of parameters in the probability functions.


\subsection{Detection Probability, $p(\bx_i, s_i)$}
\label{sec:detection_probability}

A fundamental part of ASCR is the concept of a detection probability, which is the probability that an emitted call is detected by a sensor -- this is the way ASCR accommodates for missed calls, i.e., false negatives.
The function that relates this probability to  covariates is called the detection function, denoted $p(.)$. 
For ASCR it was proposed to be most appropriate to use a detection function based on the \red{received (sound) level, also known as signal strength, which is primarily a function of source level and range} (\citealp*{Efford2009Population}; \citealp{Stevenson2015}). 
\red{We propose a detection function for sensor $j$ where $g_0 \in (0, 1)$ denotes the detection probability when the true received level of a call surpasses threshold $t_r$, and zero probability else, as follows}
\begin{equation}
    p_j(\bx_i, s_i) = g_0 \times \left(1 - \Phi \left(\frac{t_r - \E[r_{ij} | \bx_i, s_i])} {\sigma_r} \right) \right),
\end{equation}
where $\Phi$ denotes the standard normal cumulative density function (cdf) and $\sigma_r$ denotes the measurement and propagation error on received level (see Section \ref{sec:received_level}).
The threshold $t_r$ should be set by the analyst, and in this study is roughly equal to the maximum level of ocean background noise. 
\red{As there is error on the received levels, calls with an expected received level close to $t_r$ can have a detection probability between 0 and $g_0$.}

\subsection{Detected Calls, $f(n)$}
\label{sec:detected_calls}

To construct a probability function for the number of detected calls, we start with the distribution of emitted calls, which are assumed to occur independently of one another in space and time. 
Thus, let the number of emitted calls at point $\bx$ in period $T$ be a realisation of a spatially-inhomogeneous Poisson point process with intensity $D(\bx)$. 
As we only observe multiply-detected calls, we take the product of this intensity and the probability of multiply detection, denoted by 
\begin{equation}
    \begin{aligned}
        p.(\bx_i, s_i) &= \mathbb{P}(\omega^* \geq 2 | \bx_i, s_i) \\
        &= 1 - \mathbb{P}(\omega^* = 0 | \bx_i, s_i) - \mathbb{P}(\omega^* = 1 | \bx_i, s_i),
    \end{aligned} 
    \label{eqn:detected}
\end{equation}  
where $\omega^* := \sum_{j \in {1:k}} \omega_{j}$. 
Note that Equation \eqref{eqn:detected} is the part of the method that deviates from conventional ASCR and allows us to exclude all singletons.
We rewrite Equation \eqref{eqn:detected} and marginalise over source level to get the filtered Poisson point process with spatially varying rate parameter $ \int_\mathcal{S} D(\bx)p.(\bx, s)f(s)ds$.
Lastly, we marginalise over $\bx$ to get the distribution of the total number of multiply-detected calls over time $T$:
\begin{equation}
    n \sim \text{Poisson} \left(\int_{\mathcal{A}} \int_{\mathcal{S}} D(\bx) p.( \bx , s ) f(s) ds d\bx \right).
\end{equation}
The expected total number of emitted calls is then derived by integrating the density over the study area, such that $\E[N] = \int_\mathcal{A}D(\bx)d\bx$.



\subsection{Received Level,  $f(\br_i| \bw_i, \bx_i, s_i; \etab)$}
\label{sec:received_level}

Sound waves propagating through water lose strength through various mechanisms \citep{Jensen2011Computational}, a process known as `transmission loss'.  
We approximate this process by allowing a single parameter $\beta_r$ to determine the acoustic transmission loss, such that
\begin{equation}
    \E[r_{ij} | \bx_i, s_i] = s_i - \beta_r \log_{10} (d_j(\bx_i)),
\end{equation}
where $d_j(\bx_i)$ returns the distance from sensor $j$ to location $\bx_i$.
Here, $\beta_r = 20$ would reflect purely spherical spreading loss while  $\beta_r = 10$ would reflect cylindrical spreading loss; in reality the dominant process will be range- and depth-dependent, with other factors also playing a role \citep{Jensen2011Computational}.
To capture potential error in the propagation model and the measurement of received level, we follow \cite{Stevenson2015} and assume Gaussian error on the received levels, giving
\begin{equation}
    r_{ij} | \bx_i, s_i \sim \mathcal{N}\left(\E[r_{ij} | \bx_i, s_i], \sigma_r^2 \right)
\end{equation}
where $\mathcal{N}\left(\mu,\sigma^2\right)$ denotes a normal distribution with mean $\mu$ and variance $\sigma^2$. 
Unlike \cite{Stevenson2015}, we do not assume all calls above threshold $t_r$ to be detected with certainty. 
Instead, we allow for a single detection probability for calls with received levels above the threshold, since the signal processing used in our case study meant that other factors also determined detectability (see Section \ref{sec:detection_probability}). 
As $r_{ij}$ is only recorded if sensor $j$ detected the call, we condition the indexing on the $j^\text{th}$ DASAR detecting the call.
Assuming independence between the sensors, the third component of Equation \eqref{eqn:condliktractable} becomes
\begin{equation}
    \begin{aligned}
        &f(\br_i| \bw_i, \bx_i, s_i; \etab) = \\
        & \quad\quad \prod_{j \in \{1:K|\omega_{ij}=1\}} \frac{1}{\sigma_r} \frac{\phi\left( (r_{ij} - \E[r_{ij} | \bx_i, s_i])/\sigma_r \right)}{1 - \Phi (({t_r -  \E[r_{ij} | \bx_i, s_i]})/{\sigma_r} ) },
    \end{aligned}
\end{equation}
where $\phi$ denotes the standard normal probability density function (pdf).
This is in effect a normal distribution truncated at $t_r$. 

\subsection{Bearings, $f(\by_i| \bw_i, \bx_i; \gammab)$}
\label{sec:bearings}

DASARs were designed to record the direction to discrete sounds; these recorded bearings contain measurement errors. Following \cite{Stevenson2015} and \cite{Borchers2015Unifying}, we capture this error by assuming a Von Mises distribution with concentration parameter $\kappa$ on the bearings. 
Analogous to the received levels, we only record bearings at DASARs that detected a call. 
Assuming that the sensors are independent of each other, the second component of Equation \eqref{eqn:condliktractable} would therefore be
\begin{equation}
    \begin{aligned}
        &f(\by_i | \bw_i, \bx_i; \gammab) = \\
        & \quad\quad \sum_{j \in \{1:K|w_{ij} = 1\}}  \frac{\exp\{ \kappa \cos(y_{ij} - \E[y_{ij} | \bx_i]) \} }{2 \pi I_0(\kappa)},
    \end{aligned}
\end{equation}
with $I_0$ denoting the modified Bessel function of degree 0. 
Exploratory research found that, while most bearings appeared relatively accurate, a small proportion \red{seemed} highly inaccurate, and we hence apply a 2-part discrete mixture model on the bearings. 
In effect, we fit a Von Mises distribution with a lower accuracy (dispersion is $\kappa$) for some proportion of the bearings, $\psi_\kappa$, and a Von Mises distribution with higher accuracy (dispersion is $\kappa + \delta_\kappa$, where increment $\delta_\kappa$ is non-negative) to the remaining proportion of the bearings, $1 - \psi_\kappa$. 
The second component of Equation \eqref{eqn:condliktractable} thus becomes 


\begin{equation}
    \begin{aligned}
        &f(\by_i | \bw_i, \bx_i; \gammab) = \\
         & \quad\quad \sum_{j \in \{1:K|w_{ij} = 1\}}  \psi_\kappa \frac{\exp\{ \kappa \cos(y_{ij} - \E[y_{ij} | \bx_i]) \} }{2 \pi I_0(\kappa)} \\
        &\quad\quad\quad\quad\quad \times (1-\psi_\kappa)  \frac{\exp\{ (\kappa + \delta_\kappa) \cos(y_{ij} - \E[y_{ij} | \bx_i]) \} }{2 \pi I_0(\kappa + \delta_\kappa)}.
    \end{aligned}
\end{equation}

\subsection{Call location given $s$ and at least two detections, $f(\bx_i | s_i, \omega^*_i \geq 2 \red{; \phib, \thetab})$}

To evaluate the pdf of call locations, we assume independence between call locations and source levels, and use Bayes' formula to obtain
\begin{equation}
    \begin{aligned}
    f(\bx_i | s_i, \omega^*_i \geq 2 \red{; \phib, \thetab}  ) =& \frac{f( \omega^*_i \geq 2|\bx_i, s_i)f(\bx_i |s_i )}{\int_{\mathcal{A}} f( \omega^*_i \geq 2|\bx, s_i)f(\bx | s_i )d\bx} \\
    =&  \frac{p.(\bx_i, s_i )f(\bx_i)}{\int_{\mathcal{A}} p.(\bx, s_i)f(\bx)d\bx}.
    \end{aligned} \label{eqn:calllocations1}
\end{equation}
Although $f(\bx_i)$ is unknown, we do know that it is proportional to the emitted call density, such that $f(\bx_i) = {D(\bx_i)}/{ \int_{\mathcal{A}} D(\bx)d\bx}$. 
Thus we can simplify Equation \eqref{eqn:calllocations1} as
\begin{equation}
    \begin{aligned}
    f(\bx_i | s_i,  \omega^*_i \geq 2 \red{; \phib, \thetab}) =& \frac{p.(\bx_i, s_i ){D(\bx_i)}/{ \int_{\mathcal{A}} D(\bx)d\bx}}{\int_{\mathcal{A}} p.(\bx, s_i){D(\bx)}/{ \int_{\mathcal{A}} D(\boldsymbol{q})d\boldsymbol{q}}d\bx}\\ 
    =& \frac{p.(\bx_i , s_i )D(\bx_i )}{\int_{\mathcal{A}} p.(\bx, s_i  )D(\bx)d\bx}.
    \end{aligned} \label{eqn:calllocations2}
\end{equation}

\subsection{Source level given at least two detections, $f(s_i  |  \omega^*_i \geq 2 \red{; \phib, \thetab, \nub})$}
\label{sec:source_level}

Analogous to above, we assume independence between and among call locations and source levels, to derive 
\begin{equation}
    \begin{aligned}
    f(s_i  |  \omega^*_i \geq 2 \red{; \phib, \thetab, \nub} ) =& \frac{f( \omega^*_i \geq 2 | s_i  ) f(s_i  )} {f( \omega^*_i \geq 2  )} \\
    =& \frac{\int_{\mathcal{A}} p.( \bx, s_i ) D(\bx) d\bx \times f(s_i )} {\int_{\mathcal{A}} \int_{\mathcal{S}} p.(\bx, s ) f(s) D(\bx) ds d\bx}.
    \end{aligned} \label{eqn:sourcelevels1}
\end{equation}
Note that a part of the numerator in Equation \eqref{eqn:sourcelevels1} cancels out against the denominator in Equation \eqref{eqn:calllocations2}, and that the denominator of Equation \eqref{eqn:sourcelevels1} denotes the effective sampled area. 
\cite{Thode2020Roaring} estimated source levels using the estimated origin of the call, the received level on the sensor nearest the origin, and a transmission loss parameter $\beta_r$ of 15. 
Based on the observed distribution of these estimated source levels, we assume a normal distribution on $s_i$ truncated at 0, such that
\begin{equation}
    s \sim \mathcal{N}_0^{\infty}(\mu_s, \sigma_s^2).
\end{equation}

\subsection{Detection History, \red{$f(\bw_i| \bx_i, s_i, \omega^*_i \geq 2 ; \thetab)$}}
\label{sec:detection_history}

If we assume sensors to be independent, we can view the detection history of a call as a realisation of a binomial process with size $K$ and non-constant probability $p_j$ with $j = 1, ..,K$, where the order is relevant and hence the binomial coefficient is absent. This gives
\red{
\begin{equation}
    f(\bw_i | \bx_i, s_i, \omega^*_i \geq 2; \thetab) = \frac{\prod^K_{j=1}{p_j(\bx_i, s_i) ^ {\omega_{ij}} (1 - p_j(\bx_i, s_i)) ^ {1 -\omega_{ij}}} }{p.(\bx_i, s_i)},
\end{equation}
}
where $p.(\bx_i, s_i)$ appears in the denominator to account for the conditioning on at least two sensors in every call detection history (see Equation \eqref{eqn:detected}). 

\subsection{Variance Estimation}
\label{sec:variance_estimation}

We do not use the Hessian matrix to extract standard errors, as these are only asymptotically normal.
Instead, we estimate uncertainty using a non-parametric bootstrap, where we re-sample the calls with replacement \citep{Borchers2002Estimating} and fit the model every time. 
Following that, we estimate the standard error by taking the standard deviation of all bootstrapped parameter estimates, and we take the 2.5\% and 97.5\% percentiles to estimate their 95\% confidence intervals.  

\subsection{Assumptions}
\label{sec:assumptions}
The method presented in this manuscript relies on several assumptions, as follows.  
1) Call origins are a realisation of a Poisson point process, thus calls are spatially and temporally independent given this process.
2) Calls are omnidirectional and equally detectable given only the received level (i.e., no unmodelled heterogeneity). %
3) Sensors are identical in performance and operate independently;
4) Calls are matched without error and identified correctly, but can be missed (i.e., no false positive, but false negatives are allowed). 
5) The transmission loss model is correctly specified.
6) Uncertainty on bearings and received levels are independent. 
7) Source level of a call is independent of space and time.
These assumptions are discussed in more detail in Appendix B in Supplementary Materials.  

\subsection{Practical Implementation}
\label{sec:practical}
We fitted the model using maximum likelihood estimation
(MLE) in \texttt{R 4.1.0}  \citep{RCoreTeam2021} with some components written in \texttt{C++} and linked to \texttt{R} through the \texttt{Rcpp} library \citep{Eddelbuettel2013Seamless}.
We standardised the covariates in the density model to improve the convergence. The estimates were found through optimisation with the function \texttt{nlminb()} \citep{RCoreTeam2021}. 
We used a spatial mesh with non-uniform grid spacing to reduce run-time, with increased grid spacing farther from the sensor array (see  Appendix C in Supplementary Materials for details).

\section{Simulation Study}
\label{sec:sim_study}
We used simulations to evaluate the performance of the model under variable source levels (scenario 1) and fixed source levels (scenario 2). 
Both scenarios featured measurement error on bearings simulated from a two-part mixture model, and inhomogeneous call density specified as follows:
\begin{equation}
    \begin{aligned}
    \log(D) = \beta_0 + \beta_1  d + \beta_2 d^2,
    \end{aligned} \label{eqn:sim_density}
\end{equation}
where $d$ denotes the (scaled) shortest distance to the coast.  The parameter values used were chosen to match those from the case study data analysis and are given in Table \ref{table:simulationpars}. 

For each scenario, we simulated 100 data sets and analysed each data set with 5 models: (a) the true model, i.e., that corresponding to the simulated scenario; (b) a model with incorrect assumption about source level, i.e., for scenario 1 the model assumed fixed source level and for scenario 2 the model assumed variable source level; (c) a simpler bearing model assuming a von Mises distribution on bearing error but no two-part mixture; (d) a model that omitted the bearing information altogether; and (e) a model assuming a homogeneous spatial density. 
\red{We did not include a simulation scenario where we naively fit model to false positives, as the effect on the estimates is already known: it will induce a positive bias that will increase with increasing false positive rate.}

For each scenario and model combination (1a-e and 2a-e) we evaluated performance by calculating the coefficient of variation (CV), relative error and relative bias in estimated total abundance. Further details are given in Appendix D in Supplementary Materials.

\begin{table}[]
\centering
\begin{tabular}{lrr}
\hline\hline     
Parameters & Variable SL & Constant SL  \\
\hline 
$g_0$ & 0.6 & 0.6 \\
\rowcolor{lightgray}$\beta_r$ & 18.0  & 14.5   \\
\rowcolor{lightgray}$\sigma_r$ & 2.7 & 4.5 \\
$\mu_s$   & 163.0 & 155.0 \\
$\sigma_s$   & 5.0 & -  \\
\rowcolor{lightgray}$\kappa$  & 0.3 & 0.3  \\
\rowcolor{lightgray}$\delta_\kappa$  & 36.7 & 34.7 \\
\rowcolor{lightgray}$\psi_\kappa$  & 0.1 & 0.1 \\
$\beta_0$ & -12.0 & -16.0  \\
$\beta_1$ & 45.0 & 57.0   \\
$\beta_2$ & -53.0 & -68.5  \\
\hline
\end{tabular}
\caption{The parameter values used for the simulation study. These were based on initial fits to the real data, in order to keep the simulations as realistic as possible.}
\label{table:simulationpars}
\end{table}

\begin{figure} 
\centering
    \includegraphics[width=\textwidth]{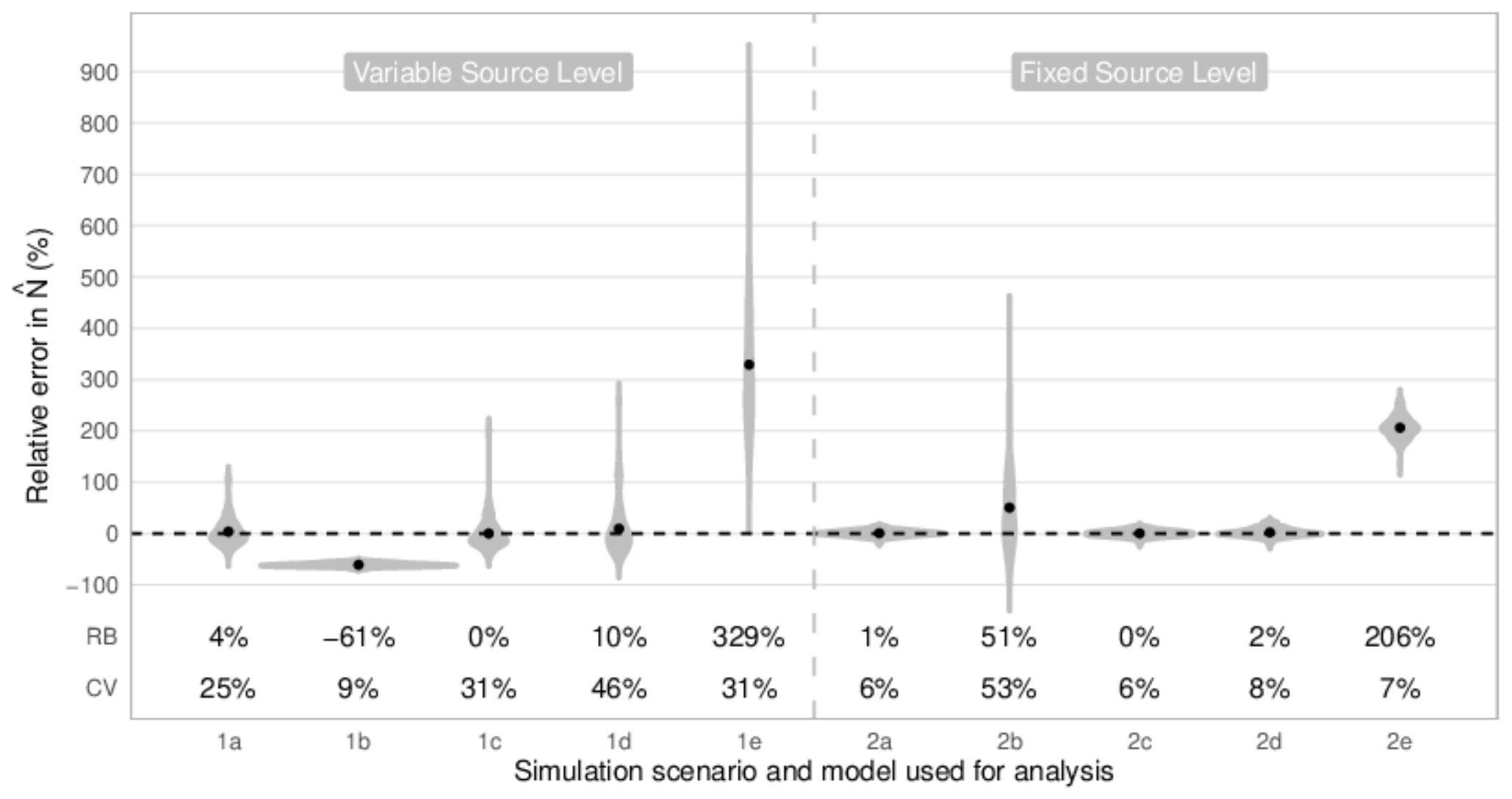}
    \caption[Relative error, relative bias and coefficient of variation for $\hat{N}$.]{Relative error, relative bias and coefficient of variation for $\hat{N}$ from 100 simulations. The black dots correspond to the relative bias (RB; the mean relative error) and the grey shading shows the spread of relative error per scenario. The RB and coefficient of variation (CV) for every scenario and model combination is shown just above the x-axis.
    Simulation scenario 1 is variable sound source level and 2 is fixed sound source level. Analysis model (a) is the true model, (b) the incorrect source level model, (c) an incorrect simple bearing model, (d) a model with bearing data, and (e) an incorrect homogeneous spatial density model. 
    }
    \label{fig:plot_model_performance}
\end{figure}

\subsection{Results}
\label{sec:sim_results}

Both correctly specified models (1a and 2a) gave near-unbiased results (Figure \ref{fig:plot_model_performance}). 
Fitting a single source level model in the variable source level scenario (1b) introduced a strong negative bias with low variance; fitting a variable source level in the fixed source level scenario (2b) introduced a strong negative bias, although this bias could be explained by a flat likelihood surface and resulting sensitivity to starting values. 
Using an incorrect, simple bearing model (1c and 2c) did not induce bias but did increase variance.
Ignoring the bearing information induced a small positive bias and greater variance in the variable source level scenario (1d) but had little effect on the fixed source level scenario (2d). 
Lastly, using an incorrectly specified constant spatial density model (1e and 2e) resulted in severe overestimation of abundance ($>300\%$ and $>200\%$ bias for variable and fixed source level scenarios, respectively).

\section{Bowhead Whale Analysis}
\label{sec:bw_analysis}

We used the proposed method to estimate bowhead whale call density and abundance at site 5 on 31 August 2010.
We used data from a single day, and as we present our estimate as call density per day, we did not need to adjust for our study time ($T = 1$).
We set $t_r$ at 96 dB, as the observed background noise never surpassed this level. 
The true call density surface is unknown and therefore we fitted several candidate models, which we compared using Akaike's information criterion (AIC; \cite{Akaike1998Information}). 
As the bowhead whales are thought to exhibit a spatial preference based on their distance from the coast and ocean depth during fall migration \citep{Greene2004Directional}, we fitted a canditate set of 35 models containing combinations of distance and depth as linear, quadratic or smooth covariates -- see Appendix E in Supplementary Materials for full details. 
Measures of uncertainty in abundance (standard error, CV, and quantile coefficient of dispersion (QCD)) were derived by bootstrapping the calls 999 times, refitting the AIC-best model each time. 
The QCD is a relative measure and insensitive to outliers, and is defined as $(Q_3 - Q_1) / (Q_3 + Q_1)$, where $Q_1$ and $Q_3$ are the first and third quartile, respectively. 
To further illustrate the potential benefits of modelling source level heterogeneity, we included point estimates of the best model equivalent with a fixed source level. 

\subsection{Results}
\label{sec:bw_results}
\begin{figure} 
    \centering
    \includegraphics[width=\textwidth]{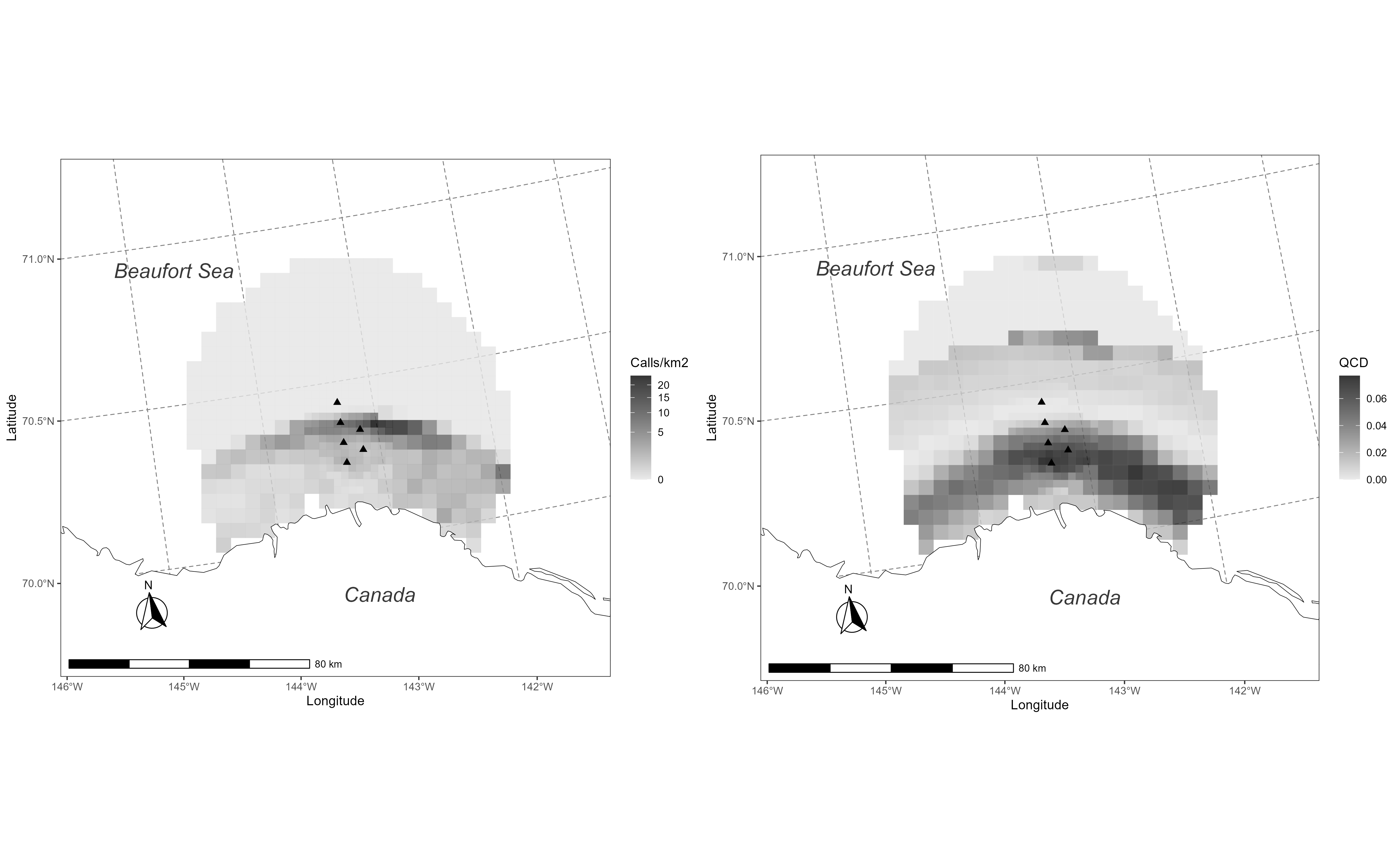}
    \caption[]{A map of the study site 5 from 31 August 2010 with the estimated density surface (left) and the associated empirical quartile coefficient of dispersion (QCD; right). Locations of the 6 DASARs are indicated with the triangles. The square grid cells within 10 km of the array each have an area of 6.25 km$^2$, and and area of 25 km$^2$ elsewhere. The density map corresponds to the model with the lowest AIC, which is model 33 in Appendix D in Supplementary Materials. The QCD is derived from 999 Monte Carlo simulations based on the same model, where the calls are re-sampled.}
    \label{fig:plot_density_qcd}
\end{figure}

The lowest AIC model included density modelled as a smooth function distance to coast and a quadratic relation with depth.
We observe an increase in density, followed by a decrease, as we increase the distance from the coast (Figure \ref{fig:plot_density_qcd}, left panel).
Moreover, density is higher just east of the sensor array, potentially due to some effect of ocean depth. 
\red{Even though we observe increases in uncertainty in the southern regions and slightly in the north, overall QCD estimates are low (Figure \ref{fig:plot_density_qcd}, right panel).}
Total abundance of bowhead whale calls was estimated at 5741 in the study area, and the CVs of the parameter estimates varied considerably, ranging from 0.95\% for $\hat{\mu}_s$ to 58.22\% for $\widehat{\beta_\text{s(depth).3}}$ (Table \ref{tab:best_model_results}). 

\begin{table}
\centering
\begin{tabular}{lrrrr} 
\toprule\textbf{Quantity}  & \textbf{Estimate} & \textbf{SE} & \textbf{CV (\%)} & \textbf{Single SL}\\
\midrule
$N$ & 5741.39 & 576.01 & 10.55 & 1916.06 \\
\\
\rowcolor{lightgray}
$g_0$ & 0.55 & 0.02 & 2.73 & 0.67 \\
$\beta_r$ & 17.05 & 0.33 & 1.91 & 13.86\\
\rowcolor{lightgray}
$\sigma_r$ & 2.75 & 0.11 & 4.06 & 4.64\\
$\mu_s$ & 158.46 & 1.50 & 0.95 & 151.04\\
\rowcolor{lightgray}
$\sigma_s$ & 5.47 & 0.26 & 4.90 & -\\
$\kappa$ & 0.77 & 0.17 & 21.95 & 0.62\\
\rowcolor{lightgray}
$\delta_\kappa$ & 45.49 & 3.02 & 6.54 & 40.23\\
$\psi_\kappa$ & 0.12 & 0.01 & 10.75 & 0.09\\
\\
\rowcolor{lightgray}
$\beta_\text{intercept}$ & -265.65 & 7.92 & 2.99 & -146.81 \\
$\beta_\text{s(depth).1}$ & 42.43 & 18.07 & 45.55 & 80.65 \\
\rowcolor{lightgray}
$\beta_\text{s(depth).2}$ & 302.26 & 6.72 & 2.23 &  8.25 \\
$\beta_\text{s(depth).3}$ & 41.14 & 22.32 & 58.22 & 88.65  \\
\rowcolor{lightgray}
$\beta_\text{s(depth).4}$ & -141.57 & 6.88 & 4.89 & -8.78  \\
$\beta_\text{s(depth).5}$ & 154.60 & 6.19 & 4.01 & 24.23 \\
\rowcolor{lightgray}
$\beta_\text{s(d2c).1}$ & 159.48 & 4.47 & 2.82 & 59.19 \\
$\beta_\text{s(d2c).2}$ & -449.58  & 15.55 & 3.47 & -185.40 \\
\rowcolor{lightgray}
$\beta_\text{s(d2c).3}$ & 101.06 & 4.29 & 4.26 & 33.88 \\
$\beta_\text{s(d2c).4}$ & -109.30 & 5.10 & 4.69 &  5.14 \\
\rowcolor{lightgray}
$\beta_\text{s(d2c).5}$ & -224.66 & 11.13 & 4.96 &  -142.24 \\
\bottomrule
\end{tabular}
\caption{\label{tab:best_model_results} The point estimates and associated uncertainties. SE is standard error, CV stands for coefficient of variation, and Single SL are point estimates from the equivalent model without source level heterogeneity. $N$ is a derived quantity; $g_0$ through $\psi_\kappa$ are the observation parameters presented on the real scale; and the remainder are the density parameters presented on their log link scale. }
\end{table}


\section{Discussion}
\label{sec:discussion}

We have developed and tested a novel extension of acoustic spatial capture-recapture by conditioning on a minimum of two detections per individual call. 
Removing single detections may be necessary when the validity of these calls is challenged, and our simulation study shows that the method gives unbiased results in both variable and single source level scenarios. 
Even though this model applies to (marine) acoustic data, the extension proposed can be applied to all forms of SCR. 
Fitting a single source level model to variable source level data, thus ignoring heterogeneity in the detectability of the calls, resulted in severe underestimation of abundance. 
This represents a cautionary tale for other applications of SCR, both terrestrial and aquatic, when it is hypothesised that there is some random variable affecting individual detectability -- it does not have to be something as tangible as a source level. 
Moreover, we confirmed results from \cite{Stevenson2015} on adding bearing information to a SCR model and found a further increase in precision when allowing for a mixture of `good' and `bad' bearings. 
Finally, we also confirm that incorrectly assuming a constant call density surface can lead to severe overestimates of the abundance. 


The notion that density of bowhead whale(s) (calls) is highest 10--75 km from the coast \citep{Greene2004Directional} seems to be confirmed by the best model, which finds a density that is highest a bit away from the coast, but not too far. 
The high call spatial density in the northeastern part of the study area (Figure \ref{fig:plot_density_qcd}) was unexpected, but could have been a consequence of using only data from one day. 
This likely strongly violated the Poisson assumption about call spatial location, since whale location is spatially auto-correlated, especially on small time scales \citep{Wursig1985Behavior}. 
Alternatively, it has been hypothesised that the higher estimated density of calls in the east could be due to the directionality of the calls or the effect of water depth on sound propagation \citep{Blackwell2012Directionality}. 
In this study, the authors observed 2.1 times as many calls east of the array as west.
Future research could explore whether a more gradual increasing/decreasing density slope is found if more days were included in the analysis, resulting in reduced autocorrelation among the call origins. 

The model-based expected number of singletons was 570 whereas there were 1417 observed singletons, which is 149\% more than expected and confirmed our suspicions regarding false positives (see Figure \ref{fig:percentage_dasars_involved}). 

\red{An alternative to truncating singletons would be to retain all data and try to model the occurrence of false positive detections.
For example, one could assume that an unknown proportion of detections are false positives and include a parameter for this proportion in an analogous way to the $M_{t, \alpha}$ model presented by \cite{Yoshizaki2007Use}.
Detection at any detector $j$ could have three outcomes: no detection with probability $1-p_j$, false detection with probability $\alpha p_j$, and correct detection with probability $p_j$. 
Moreover, we would have to implement some modification of the $M_b$ model \citep{Otis1978Statistical} to account for a change in $\alpha$ for the multiply-detected calls, which would likely involve an additional parameter.
A benefit of this $M_{b, \alpha}$ model would be that it could allow for false positives in multiply-detected calls, however, it increases complexity and introduces hard-to-test assumptions on the false positive rate.
Given that a large amount of acoustic data are potentially available (we used data from just one day), including poor-quality data seemed undesirable given the additional complexity and run time.
Hence we consider the truncation approach developed here to be best for our application of ASCR.}

The main model in our study conditions on at least two positive detections for every detection history, but can readily be extended to condition on a minimum of three or more involved sensors. 
This requires generalising Equation \ref{eqn:detected} to
\begin{equation}
    \begin{aligned}
        p.(\bx, s) =& \mathbb{P}(\omega^* \geq \omega^*_\text{min} | \bx, s) \\
        =& 1 - \mathbb{P}(\omega^* = 0 | \bx, s) - \mathbb{P}(\omega^* = 1 | \bx, s) \\
        &\quad- ... - \mathbb{P}(\omega^* = \omega^*_\text{min} - 1 | \bx, s),
    \end{aligned} 
    \label{eqn:detected_general}
\end{equation} 
where $\omega^*_\text{min}$ denotes the minimum number of sensors involved.
Preliminary simulations showed no apparent bias, although depending on the situation, conditioning this way can rapidly reduce the amount of data available and hence decrease estimator precision. This is illustrated by the relative frequencies in Figure \ref{fig:percentage_dasars_involved}.

We estimated variance empirically by bootstrapping the calls. 
This assumes independence between the calls, which we know is violated as the calls are produced by whales and whales are spatially autocorrelated. 
Moreover, calls can potentially trigger responses from nearby whales, leading to temporal correlation between them \citep{Thode2020Roaring}. 
Some of the spatial and temporal dependence was eliminated by removing all detections with a \red{received level} below 96 dB, as this thinned the data, assuming independence between call characteristics and ocean noise. 
To create more accurate variance estimates, one could \red{stratify} the data by time and bootstrap these time chunks, thereby removing some of the autocorrelation. 
Even better, one could sample short time frames from several days and combine those to obtain the desired amount of data to analyse, in which case the analysed observations themselves could be assumed independent and likelihood-based variance estimates are acceptable (given a large enough data set). 
The length and frequency of these time chunks will be case specific, as this will depend on the calling behaviour of the studied species.

We estimated model parameters using MLE as opposed to in a Bayesian inferential framework, with MLE having coding simplicity and reduced run time as the main benefits.
We did find that convergence was sometimes slowed due to flat likelihood surfaces as a consequence of the many sources of variation in our model. 
A benefit of using a Bayesian framework would be the possibility to include informative prior distributions based on previous research, especially on the latent variable source level, which would likely improve convergence and may improve precision of density and abundance estimates. 


Our method derives a call abundance, which can be converted into an animal abundance by correcting this estimate for the call rate, similarly to \cite{Marques2013}. 
Ideally, this call rate is estimated for the same population and alongside the collection of the data, but this is not always possible.
\cite{Blackwell2021} estimated a call rate for the BCB population over a longer period of time. 
Naively dividing our estimate or total calls by their median call rate estimate of 31.2 calls/whale/day (interquartile range of 12-129.6 calls/whale/day)  gives us $\widehat{N}_\text{whales} = 5741.39 / 31.2 \approx 184$ individuals.
We present this number for illustration alone; for correct inference, one should properly account for the variance around call rate and call abundance estimates.
Alternatively, there are methods that directly estimate animal density based on cues (for more information, see e.g., \cite{Fewster2016TraceContrast} and \cite{Stevenson2019Cluster}).

When background noise is highly variable, selecting a truncation level that ensures that noise (almost) never surpasses the received level might result in most data being discarded. 
When data are scarce, it may then be beneficial to use a detection function based on the signal-to-noise ratio (SNR).
Here, detection probability is assumed to depend on the ratio of the received level to the background noise, and could increase either step-wise or gradually as a function of SNR. 
This way, no data will be lost due to truncation.
However, using an SNR detection function requires a random sample of accurate noise measurements for an additional Monte Carlo integration in the likelihood, and noise measurements at all sensors for every cue with a detection history to be able to estimate the detection probabilities.
A detailed description of the SNR likelihood is presented in Appendix F in Supplementary Materials. 

A potential weakness of the model is the fact that we assume the same propagation loss model for the entire study area. 
For most of the study site this assumption is reasonable, as it concerns a shallow plateau with little variation in ocean depth. 
However, the ocean floor drops in the northern part of the study site, resulting in altered propagation conditions.
It is assumed that the bowhead whales migrate mainly over the shallow plateau and hence depth-induced inhomogeneous propagation is not a practical issue in our case study. 
In general, however, it is something that should be considered when modelling sound propagation in variable (ocean) landscapes.
\cite{Phillips2016Passive} and \cite{Royle2018Modelling} presented ways in which it is possible explicitly to model variable sound attenuation. 
Such a model requires accurate and sometimes high resolution information on environmental variables that affect sound attenuation, such as ocean depth.

The model presented here extends the scope of SCR to provide reliable inference on spatial density and abundance from passive acoustic data.  
Removing single detections will also be of use in other SCR applications where single detections are unreliable -- for example when association between detections is not perfect such that repeat detections of the same individual are sometimes incorrectly classified as single detections of a new individual.  
This can happen in situations where natural animal characteristics are used for identification, such as photo- and genetic-ID. 

\section*{Acknowledgements}


\section*{Declarations}

\textbf{Code and data availability} The code to run the simulation, and the code and data used to fit the models, are publicly available at https://github.com/fpetersma/bowhead\_whale\_ASCR.

\noindent\textbf{Conflict of interest} The authors declare that they have no conflict of interest.

\bibliography{references}







\label{lastpage}

\end{document}


\maketitle

\tableofcontents

\newpage

\section{Appendix A: Data Background, Availability \& Cleaning}\label{appendix-a-data-background-availability-cleaning}

\subsection{Data availability}\label{data-availability}

To handle the vast number of calls detected by the DASARs, \cite{Thode2012aAutomated} developed a method to detect and match calls using a neural
network trained using manually processed data from 2008 and 2009. From
these data a collection of detection histories could be derived, where
for every call and every DASAR it contains a 1 if detected, and a 0
otherwise. Sometimes, calls would be wrongly identified and/or matched,
as other discrete sounds produced by, among others, airguns, bearded
seals (\textit{Erignathus barbatus}) and walruses (\emph{Odobensu
rosmarus divergens}) can appear similar to bowhead whale calls \citep{Thode2012aAutomated}. It was thought that single detection calls (`singletons')
contain a high proportion of falsely identified calls (false positives),
thus we removed these detections altogether.

In this study, we focus on the data from one specific day, 31 August
2010, from site 5, the most eastwards site, as this was of one the more
intensely studied subsets of the data. Moreover, calls accumulated
somewhat evenly across the day, resulting in a low expected rate of
overlapping calls. For this day, the following observations were
extracted:

\begin{itemize}
\item
  A matrix containing detection histories for all calls and all sensors,
  with 1 indicating detections and 0 indicating no detection.
\item
  A matrix containing received levels for the positive detections, and
  NA values otherwise.
\item
  A matrix containing noise measurements for all sensors and all calls.
  Noise at DASARs that detected a call was derived using a 6 second
  window, consisting of 3 seconds before and 3 seconds after the call,
  using the same frequency band as the call. Noise recordings for DASARs
  that did \textit{not} detect a call were derived using the widest
  observed frequency window for the DASARs that were involved in the
  detection of the call, using a similar time window.
\item
  A matrix containing noise measurements derived over 6-second windows,
  taken at a random time points and frequency bands within the study
  period. These were subsequently checked to not contain any calls, as
  that would not qualify as `noise' - other discrete sounds, such as
  walrus calls or airguns, do qualify as noise, as they potentially
  interfere with the identification of bowhead whale calls.
\end{itemize}

The last two matrices containing noise information are only used for the
detection function based on SNR (see Appendix F).

Two variables in our likelihood are latent: the origin of the call and
the source level of the call. To limit the run time, we only integrate
over the smallest subset of the domain that includes at least 99.9\% of
the associated probability density/surface, at the lowest resolution
that ensures good estimates. Thus, the remainder of our data includes:

\begin{itemize}
\item
  A matrix of at least 300 evenly spaced spatial points covering an area
  such that calls that originated at the bounds have a probability of
  being detected on at least two sensors of \(< 0.1%
  \). For every grid point, the ocean depth and distance to coast was
  also known; these were used as potential covariates in the density
  model.
\item
  A vector of source levels, ranging from 100 to 220 dB, with increments
  of 3 dB, which were the largest increments that still resulted in good
  parameter estimates.
\end{itemize}

\subsection{Data cleaning}\label{data-cleaning}

ASCR theory requires perfect identification, and to meet this assumption
as closely as possible, several cleaning procedures were required. The
false positives among calls that involved two or more detections did not pose a problem for the original studies \citep{Thode2012aAutomated}. Here, the main goal was to triangulate the bearings of detected
calls to estimate the spatial origin of the call. Every involved DASAR
would record a bearing, and a localisation procedure would add varying
weights to these bearings until it resulted in a successful
localisation, or not \citep{Thode2012aAutomated}. For example, if the bearings
of a call with two detections would never meet, e.g., because one of
them belonged to a wrongly identified sound, a localisation would fail.
If, however, a call that involved three detections had two bearings that
matched and one that did not (the `outlier'), it would give the latter a
weight close to 0, and the others a weight closer to 1, thus resulting
in a successful localisation. As we only wanted the `valid' detections
in our detection histories, we used these weights to remove wrongly
identified and matched calls. This was done using the following
procedure:
\red{
\begin{enumerate}
\item We start by only including call events that led to a successful localisation. This first step got rid of misidentified airgun signals, as those originated too far from to array to be localisable. 
\item
  If the call event contained at least one bearing with a weight $<  0.02$,
  we removed the detection with the lowest weight.
\item
  The call was localised again (now excluding the previously removed bearing), and the newly assigned weights were evaluated.
\item
  If the call was successfully localised and still contained at least
  one weight $< 0.02$, we removed that detection and went back to step
  2. Otherwise, the call was now considered valid.
\end{enumerate}
}
The weight of 0.02 was found to most accurately remove wrongly
identified and matched calls; through visual checks we found that lower values still allowed for some incorrect bearings to be included, and that increase the weight did not make a difference in how many incorrect bearings were removed. 
Finally, one call had a failed noise recording and was thus excluded from the analysis.

This left us with a total of 5793 \red{`correct'} calls detected on at least two sensors
at site 5 on 31 August 2010. The next stage was to truncate the data to
include only detections that exceed the highest sampled background noise
level of 96dB.\\
This left 443 calls after truncation that were used in the analysis
reported in the main paper.

\subsection{\red{(In)accurate bearings}}

\red{When visually exploring our data, we noticed that not all bearings were equally precise. 
Plotting a subset of the bearings suggested a split in bearings of mostly accurate and some inaccurate ones. 
It was there hypothesised to best allow for a two-part mixture model on bearing accuracy, as presented in the main paper.
An example of inaccurate versus accurate bearings is presented in Figure \ref{fig:bearing_plots} below.}

\begin{figure}
    \centering
    \includegraphics[width=\textwidth]{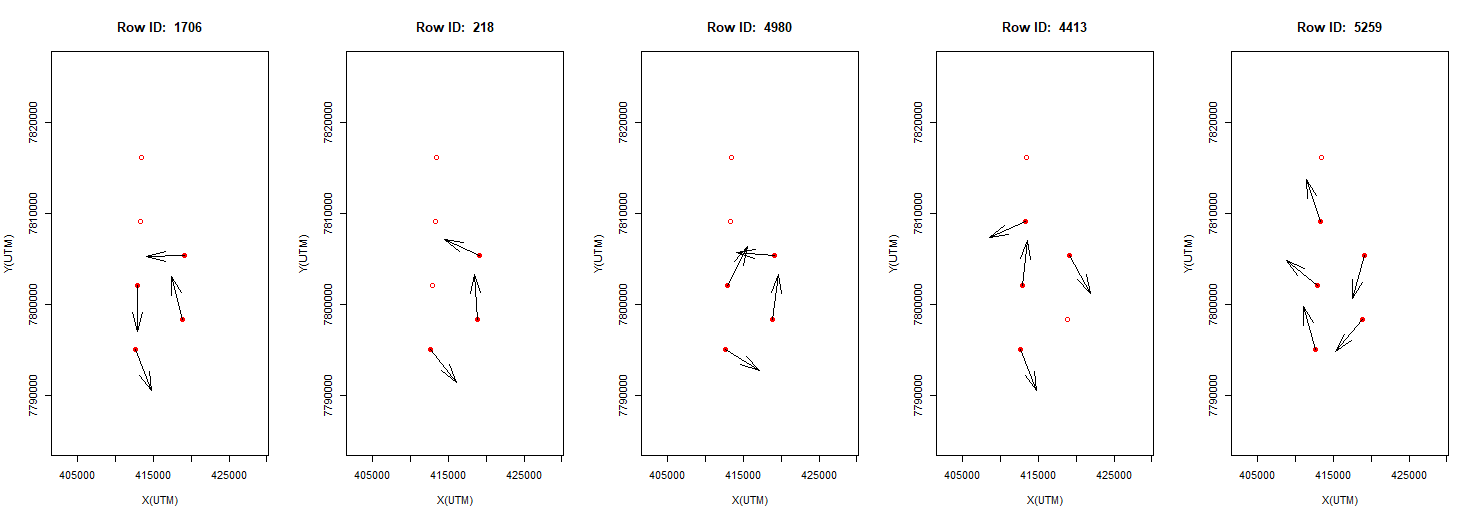}
    \caption{Plots of five calls from the data from Site 5 on 31 August 2020. Most bearings are fairly accurate, but these five plots show that can occasionally be way off.}
    \label{fig:bearing_plots}
\end{figure}

\newpage

\section{Appendix B: Model assumptions}

An overview of the model assumptions is presented below.

\begin{enumerate}
\item
  \textbf{Call origins are a realisation of a Poisson point process,
  thus calls are spatially and temporally independent.} Because of this,
  we assume that the number of calls is a Poisson random variable. The
  assumption is likely to be violated due to the spatial dependence of a
  whale (when it calls several times) and the theorised communicative
  nature of the calls. Violation likely mainly affects uncertainty
  estimates, not point estimates. Thinning the data could ensure that
  this assumption is closer to being met. Alternatively, one could
  bootstrap by time blocks to ensure more robust variance estimates.
\item
  \textbf{Calls are omnidirectional and equally detectable given only
  the received level (i.e., no unmodelled heterogeneity).} It is very
  likely that other factors impact the detectability of these calls,
  such as the frequency and duration. The direction and severity of a
  potential bias due to an incorrectly specified detection function is
  not known.
\item
  \textbf{Sensors are identical in performance and operate
  independently.} This means that we can model a single detection
  probability.
\item
  \textbf{Calls are matched without error and identified correctly, but
  can be missed (i.e., no false positives, but false negatives are
  allowed).} This is an essential assumption to the model, as we only
  allow for false negatives through the concept of a detection
  probability, but not for false positives. Wrongly identified calls
  will lead to an overestimate of abundance. Incorrectly matched calls
  will tend to lead to an underestimate of abundance.
\item
  \textbf{The transmission loss model is correctly specified.} Our
  propagation model is relatively simple and assumes a single
  transmission loss parameter, and is surely an oversimplification. A
  more complex (and realistic) propagation model could be used,
  potentially calculating propagation loss in the target frequency band
  of calls between all grid points and all sensors, and then using this
  as a look-up table in a non-Euclidian SCR framework \citep{Royle2018Modelling, Phillips2016Passive}. For the current case study, we believe this would make
  relatively little difference to the results since bathymetry is the
  main driver of variation in propagation loss in this area and almost
  all calls are believed to have originated from the same shallow shelf
  area as the sensors. However, this a topic worthy of further
  investigation.
\item
  \textbf{Uncertainty on bearings and received levels are independent.}
  One way this assumption could be violated is if there are certain
  characteristics of the water column that affect both the precision of
  the bearing measurements and the received level measurements. This
  will likely not affect the point estimates.
\item
  \textbf{Source level is independent of space and time.} This
  assumption could be violated for a variety of reasons. Firstly, if
  calls are communicative in nature, then potentially this could make
  the loudness depend on space and time. Some whales could also be
  louder than others, which means that the source level of a call is
  dependent on space and time through the individuals who produced it.
  Moreover, calls could be louder when ocean background noise is louder,
  an effect that is also known as the `Lombard effect' \citep{Thode2020Roaring}. When ocean noise is then spatio-temporally independent, so is
  source level. We believe that a violation of this assumption would
  mainly affect uncertainty estimates and not point estimates.
\end{enumerate}

\newpage

\section{Appendix C: Spatial Mesh}

In spatial capture-recapture, and therefore also in acoustic spatial
capture-recapture, we estimate a density surface. Total abundance is
derived by integrating over space, giving a total abundance for the
area. In practice, we approximate the integral by summing the function
over a discrete mesh. We cannot compute the approximation over infinite
2-dimensional space, so we only approximate the integral over our study
area \(A \subset R^2\) The condition on this study area is that the
probability of a call originating at the bounds or farther out has a
probability of being detected on at least two sensors is at most
\(1\%\). This makes the integration computationally feasible whilst
inducing negligible bias.

In our case study, loud calls could potentially be audible hundreds of
kilometres from the source. To keep computer run-time low, we had to
limit the number of grid cells in the order of 500, which limits the
resolution of our mesh. As our study site did not contain many steep
gradients in the covariates, we did not need a high resolution to
capture this. However, we did need a finer resolution closer to the
sensor array to accommodate the bearings. Recall that we included
bearing data with errors in our model. If we used a low resolution grid
close to our array, we would have greatly reduced the probability that
one of the midpoints of our grid cells was close to what the recorded
bearings point towards, and thus it would have been hard for the model
to estimate the bearing accuracy. Our solution was to use a mesh with
increasing grid spacing the farther the cells are from the sensor array.

This mesh had a two-stage grid spacing: the `high resolution' cells
within a 10,000 meter radius from the outer sensors of the array had an
area of \(6.25 \text{ km}^2\) each; the `low resolution' cells between
10,000 and 50,000 meters from the outer sensors had an area of
\(25 \text{ km}^2\). This way we created higher resolution near the
sensors whilst keeping the number of grid cells in the mesh to 438.
Every grid cell was rectangular on the Albers projection, and the
associated covariate information (i.e., ocean depth and distance to the
coast) was derived for the midpoint of the cell. Plots of the mesh with
detection probabilities associated with the simulation parameters (which
were based on parameter estimates from the real data) are presented in
Appendix D. These plots that the detection probabilities are well below
\(1\%\) at the bounds.
 
\section{Appendix D: Simulation
Study}\label{appendix-d-simulation-study}

Simulation studies were performed to assess the general performance of
the implemented method. Data were simulated in the following way:

\begin{enumerate}
\item
  Derive the expected density for every grid point and simulate the
  number of emitted calls for every mesh cell \(m \in 1,...,M\) from
  \(\text{Poisson}(\lambda = a_m D_m)\), where \(a_m\) and \(D_m\) are
  area and density per unit area for cell \(m\), respectively.
\item
  Ensemble a dataset of all emitted calls with all their characteristic
  information.
\item
  Calculate the distances for the calls to all sensors.
\item
  For all calls, derive the bearings from the sensors to their
  locations, and add error from \(\text{VonMises}(0, \kappa)\).
\item
  For every call, simulate a source level from a zero-truncated Normal
  distribution \(\mathcal{N}^\infty_0(\mu_s, \sigma_s^2)\).
\item
  Using the distances and the transmission loss parameter \(\beta_r\),
  derive the received levels for all calls at all sensors, and add
  measurement error from \(\mathcal{N}(0, \sigma_r^2)\).
\item
  Assign detection probability \(g_0\) to every call at a sensor when
  \(r_{ij}\) is at least truncation value \(t_r\), and 0 otherwise.
\item
  Simulate from \(U(0,1)\) and accept as a positive detection if the
  value simulated is less than or equal to the detection probability.
  Then, remove all calls and associated data if the total number
  detections for that call is less than two.
\item
  Return the detection histories, received levels, and bearings for the
  remaining calls.
\end{enumerate}

Alternatively, if a mixture on the bearing precision is desired, step
(4) is replaced by:

4a. For all calls, derive the bearings from the sensors to the call. 4b.
Simulate from \(U(0,1)\) and label the detection of a call on a sensor
as `bad/low-precision' if the value simulated is less than or equal to
(\(\psi_\kappa\)). To the bearings for these sensor-call interactions,
add error from VonMises(0, \(\kappa\)). 4c. For the remaining bearings
(the `good/high-precision' bearings) add an error from VonMises(0,
\(\kappa + \delta_\kappa\)). We link the two distributions through their
dispersion parameters to ensure identifiability by making it explicit
that the second distribution always has a larger dispersion parameter,
and thus lower variance.

\subsection{Simulation studies for functionality}\label{simulation-studies-for-functionality}

We first performed several simulation studies to evaluate the bias and
variance of the estimates. The code used to simulate data can be found
in \texttt{simulate\_data\_Rcpp.R}. The parameters values used for the
simulations were based on exploratory fits on the real data. Using a
truncation value of 96 dB for the detection function, we fit a
homogeneous and non-homogeneous density model. Both models included a
mixture on bearing precision and source level as a latent variable.

We did not centre the covariates because this makes the spatial density
covariates harder to interpret, and an exploratory analysis with centred
covariates yielded almost identical results.

\begin{longtable}[]{@{}
  >{\raggedleft\arraybackslash}p{(\columnwidth - 6\tabcolsep) * \real{0.3611}}
  >{\raggedleft\arraybackslash}p{(\columnwidth - 6\tabcolsep) * \real{0.2361}}
  >{\raggedleft\arraybackslash}p{(\columnwidth - 6\tabcolsep) * \real{0.2222}}
  >{\raggedleft\arraybackslash}p{(\columnwidth - 6\tabcolsep) * \real{0.1806}}@{}}
\caption{Parameter estimates for the variable source level model with a
homogeneous or inhomogeneous density model, and the simulation
parameters.}\tabularnewline
\toprule
\begin{minipage}[b]{\linewidth}\raggedleft
Parameter
\end{minipage} & \begin{minipage}[b]{\linewidth}\raggedleft
Homogeneous density
\end{minipage} & \begin{minipage}[b]{\linewidth}\raggedleft
Inhomogeneous density
\end{minipage} & \begin{minipage}[b]{\linewidth}\raggedleft
Simulation
\end{minipage} \\
\midrule
\endfirsthead
\toprule
\begin{minipage}[b]{\linewidth}\raggedleft
Parameter
\end{minipage} & \begin{minipage}[b]{\linewidth}\raggedleft
Homogeneous density
\end{minipage} & \begin{minipage}[b]{\linewidth}\raggedleft
Inhomogeneous density
\end{minipage} & \begin{minipage}[b]{\linewidth}\raggedleft
Simulation
\end{minipage} \\
\midrule
\endhead
\(g_0\) & 0.53164 & 0.55022 & 0.6 \\
\(\beta_r\) & 19.14684 & 17.65528 & 18.0 \\
\(\sigma_r\) & 2.62344 & 2.70003 & 2.7 \\
\(\mu_s\) & 166.62167 & 162.46326 & 163.0 \\
\(\sigma_s\) & 5.13998 & 4.98272 & 5.0 \\
\(\kappa\) & 0.30208 & 0.23774 & 0.3 \\
\(\delta_\kappa\) & 36.94962 & 37.45905 & 36.7 \\
\(\psi_\kappa\) & 0.10337 & 0.09690 & 0.1 \\
\(\beta_0\) & 1.20156 & -12.08586 & -12.0 \\
\(\beta_\text{d}\) & - & 45.48018 & 45.0 \\
\(\beta_{\text{d}^2}\) & - & -53.83861 & -53.0 \\
\bottomrule
\end{longtable}

Our second simulation scenario involved simulating from a model with a
fixed source level. Fitting a fixed source level model to the data, we
get the following parameter estimates.

\begin{longtable}[]{@{}
  >{\raggedleft\arraybackslash}p{(\columnwidth - 6\tabcolsep) * \real{0.3611}}
  >{\raggedleft\arraybackslash}p{(\columnwidth - 6\tabcolsep) * \real{0.2361}}
  >{\raggedleft\arraybackslash}p{(\columnwidth - 6\tabcolsep) * \real{0.2222}}
  >{\raggedleft\arraybackslash}p{(\columnwidth - 6\tabcolsep) * \real{0.1806}}@{}}
\caption{Parameter estimates for the fixed source level model with a
homogeneous or inhomogeneous density model, and the simulation
parameters.}\tabularnewline
\toprule
\begin{minipage}[b]{\linewidth}\raggedleft
Parameter
\end{minipage} & \begin{minipage}[b]{\linewidth}\raggedleft
Homogeneous density
\end{minipage} & \begin{minipage}[b]{\linewidth}\raggedleft
Inhomogeneous density
\end{minipage} & \begin{minipage}[b]{\linewidth}\raggedleft
Simulation
\end{minipage} \\
\midrule
\endfirsthead
\toprule
\begin{minipage}[b]{\linewidth}\raggedleft
Parameter
\end{minipage} & \begin{minipage}[b]{\linewidth}\raggedleft
Homogeneous density
\end{minipage} & \begin{minipage}[b]{\linewidth}\raggedleft
Inhomogeneous density
\end{minipage} & \begin{minipage}[b]{\linewidth}\raggedleft
Simulation
\end{minipage} \\
\midrule
\endhead
\(g_0\) & 0.61287 & 0.65202 & 0.6 \\
\(\beta_r\) & 17.00578 & 14.54507 & 14.5 \\
\(\sigma_r\) & 4.20759 & 4.51637 & 4.5 \\
\(\mu_s\) & 163.73203 & 153.98830 & 155.0 \\
\(\kappa\) & 0.40459 & 0.31278 & 0.3 \\
\(\delta_\kappa\) & 33.16735 & 35.88457 & 34.7 \\
\(\psi_\kappa\) & 0.11720 & 0.10118 & 0.1 \\
\(\beta_0\) & 0.01268 & -16.09090 & -16.0 \\
\(\beta_\text{d}\) & - & 56.95405 & 57.0 \\
\(\beta_{\text{d}^2}\) & - & -68.37234 & -68.5 \\
\bottomrule
\end{longtable}

\subsection{Checking the buffer
width}\label{checking-the-buffer-width}

Figures 1--4 are associated with the parameters from Table 1,
corresponding to the model with a variable source level. Figures 5--8
are associated with the parameters from Table 2, corresponding to the
model with a fixed source level. They show the midpoints of the grid
cells, and whether a call produced at that point is expected to surpass
the multiply-detection threshold (blue) or not (pink), which is either
\(1%
\) or \(0.1%
\). We used a threshold of \(0.1%
\) for our study, but these figures show how we could have reduced the
size or our mesh by increasing this threshold to \(1%
\).

\begin{figure}
\includegraphics[width=0.8\linewidth]{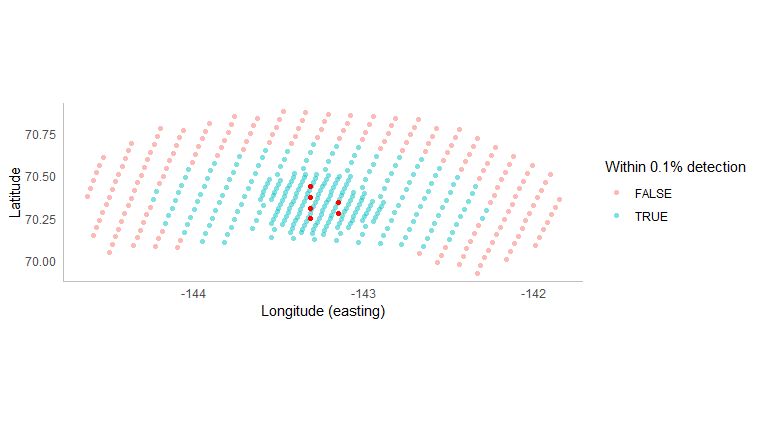} \caption{Plot of midpoints of the grid cells. Blue indicates that the expected probability of detecting a call produced at that point at least twice is at least the threshold, and pink indicates otherwise. Detection probabilities were derived from the homogeneous density parameters for the variable source level model. The sensor array is indicated in red.}\label{fig:unnamed-chunk-1}
\end{figure}

\begin{figure}
\includegraphics[width=0.8\linewidth]{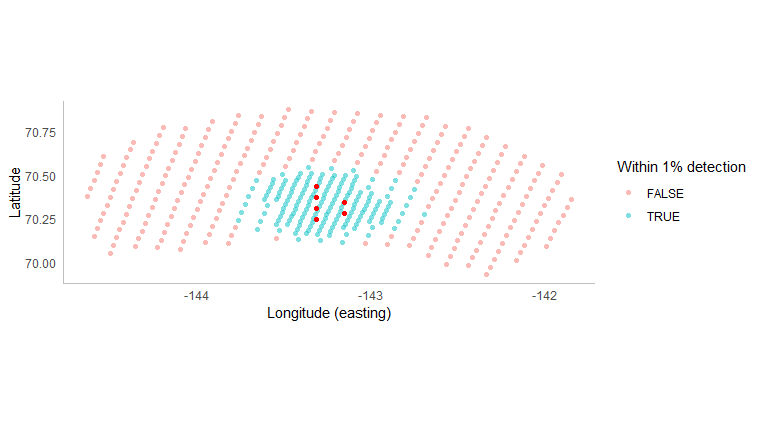} \caption{Plot of midpoints of the grid cells. Blue indicates that the expected probability of detecting a call produced at that point at least twice is at least the threshold, and pink indicates otherwise. Detection probabilities were derived from the homogeneous density parameters for the variable source level model. The sensor array is indicated in red.}\label{fig:unnamed-chunk-2}
\end{figure}

\begin{figure}
\includegraphics[width=0.8\linewidth]{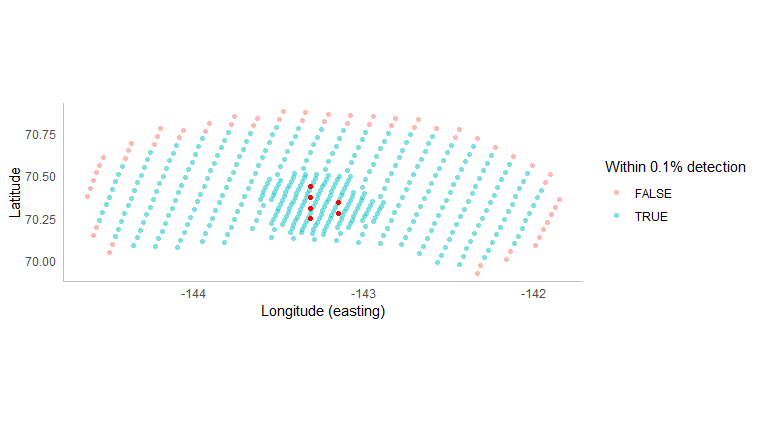} \caption{Plot of midpoints of the grid cells. Blue indicates that the expected probability of detecting a call produced at that point at least twice is at least the threshold, and pink indicates otherwise. Detection probabilities were derived from the inhomogeneous density parameters for the variable source level model. The sensor array is indicated in red.}\label{fig:unnamed-chunk-3}
\end{figure}

\begin{figure}
\includegraphics[width=0.8\linewidth]{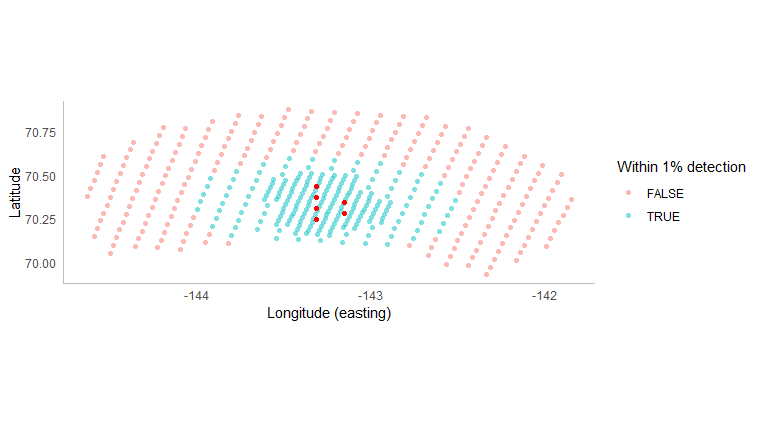} \caption{Plot of midpoints of the grid cells. Blue indicates that the expected probability of detecting a call produced at that point at least twice is at least the threshold, and pink indicates otherwise. Detection probabilities were derived from the inhomogeneous density parameters for the variable source level model. The sensor array is indicated in red.}\label{fig:unnamed-chunk-4}
\end{figure}

The plots below are associated with the parameters from Table 2,
corresponding to models with a fixed source level. They show the
midpoints of the grid cells, and whether a call produced at that point
is expected to surpass the multiply-detection threshold (blue) or not
(red), which is either \(1%
\) or \(0.1%
\). The red points represent the sensor array.

\begin{figure}
\includegraphics[width=0.8\linewidth]{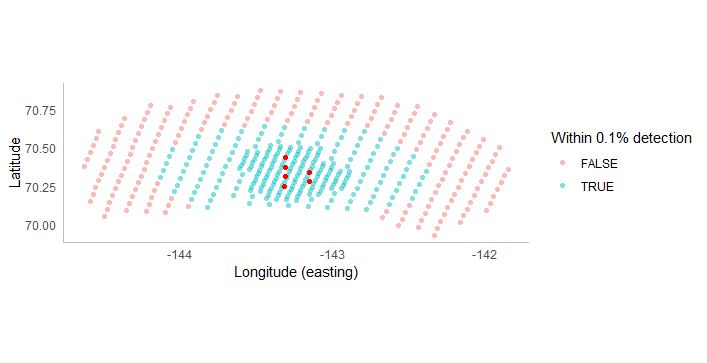} \caption{Plot of midpoints of the grid cells. Blue indicates that the expected probability of detecting a call produced at that point at least twice is at least the threshold, and pink indicates otherwise. Detection probabilities were derived from the homogeneous density parameters for the fixed source level model. The sensor array is indicated in red.}\label{fig:unnamed-chunk-5}
\end{figure}

\begin{figure}
\includegraphics[width=0.8\linewidth]{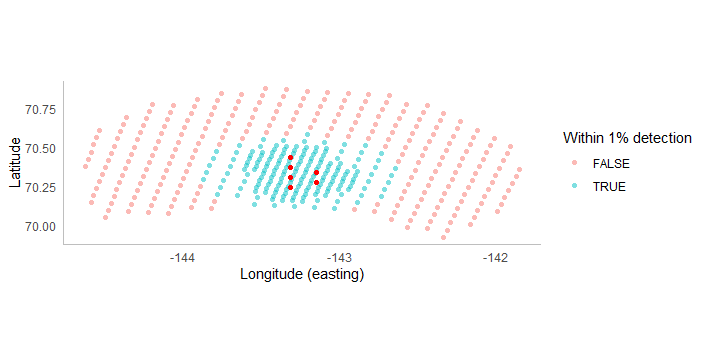} \caption{Plot of midpoints of the grid cells. Blue indicates that the expected probability of detecting a call produced at that point at least twice is at least the threshold, and pink indicates otherwise. Detection probabilities were derived from the homogeneous density parameters for the fixed source level model. The sensor array is indicated in red.}\label{fig:unnamed-chunk-6}
\end{figure}

\begin{figure}
\includegraphics[width=0.8\linewidth]{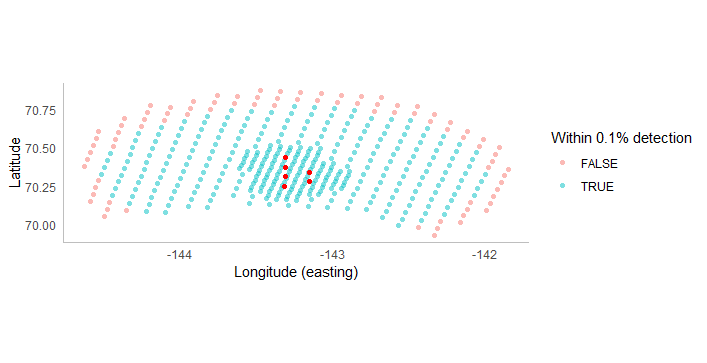} \caption{Plot of midpoints of the grid cells. Blue indicates that the expected probability of detecting a call produced at that point at least twice is at least the threshold, and pink indicates otherwise. Detection probabilities were derived from the inhomogeneous density parameters for the fixed source level model. The sensor array is indicated in red.}\label{fig:unnamed-chunk-7}
\end{figure}

\begin{figure}
\includegraphics[width=0.8\linewidth]{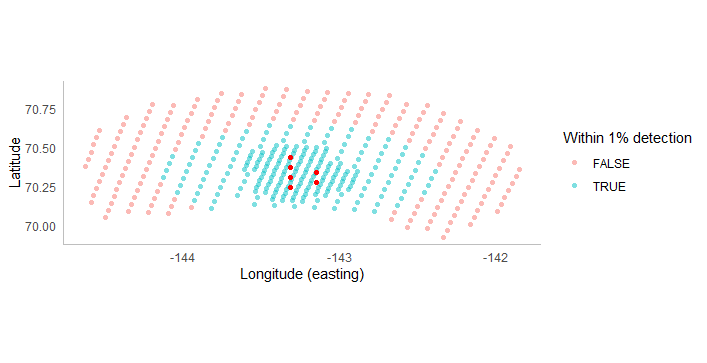} \caption{Plot of midpoints of the grid cells. Blue indicates that the expected probability of detecting a call produced at that point at least twice is at least the threshold, and pink indicates otherwise. Detection probabilities were derived from the inhomogeneous density parameters for the fixed source level model. The sensor array is indicated in red.}\label{fig:unnamed-chunk-8}
\end{figure}

\newpage

\section{Appendix E: Candidate Density
Models}\label{appendix-e-candidate-density-models}

In our study, we fitted a wide range of models to the data from our case
study. 35 models were tried, including a naive constant density model,
for completeness. Earlier studies found that the bowhead whales tend to
migrate between 10 and 75 kilometres from the coast, over a plateau that
is only between 10 and 35 meters deep. Thus, we limited our models to
combinations of those variables, with and without interaction.

Below we present the models that were tried, as an R code chunk:

\begin{Shaded}
\begin{Highlighting}[]
\DocumentationTok{\#\# The candidate density models}
\NormalTok{models }\OtherTok{\textless{}{-}} \FunctionTok{list}\NormalTok{(D }\SpecialCharTok{\textasciitilde{}} \DecValTok{1}\NormalTok{,}
\NormalTok{               D }\SpecialCharTok{\textasciitilde{}}\NormalTok{ distance\_to\_coast }\SpecialCharTok{+}\NormalTok{ distance\_to\_coast2,}
\NormalTok{               D }\SpecialCharTok{\textasciitilde{}}\NormalTok{ distance\_to\_coast }\SpecialCharTok{+}\NormalTok{ distance\_to\_coast2 }\SpecialCharTok{+}\NormalTok{ distance\_to\_coast3,}
\NormalTok{               D }\SpecialCharTok{\textasciitilde{}}\NormalTok{ depth,}
\NormalTok{               D }\SpecialCharTok{\textasciitilde{}}\NormalTok{ depth }\SpecialCharTok{+}\NormalTok{ depth2,}
\NormalTok{               D }\SpecialCharTok{\textasciitilde{}}\NormalTok{ logdepth,}
\NormalTok{               D }\SpecialCharTok{\textasciitilde{}}\NormalTok{ logdepth }\SpecialCharTok{+}\NormalTok{ depth }\SpecialCharTok{+}\NormalTok{ depth2,}
\NormalTok{               D }\SpecialCharTok{\textasciitilde{}}\NormalTok{ logdepth }\SpecialCharTok{+}\NormalTok{ distance\_to\_coast }\SpecialCharTok{+}\NormalTok{ distance\_to\_coast2 }\SpecialCharTok{+} 
\NormalTok{                 distance\_to\_coast3,}
\NormalTok{               D }\SpecialCharTok{\textasciitilde{}}\NormalTok{ logdepth }\SpecialCharTok{+}\NormalTok{ depth }\SpecialCharTok{+}\NormalTok{ distance\_to\_coast }\SpecialCharTok{+}\NormalTok{ distance\_to\_coast2,}
\NormalTok{               D }\SpecialCharTok{\textasciitilde{}}\NormalTok{ logdepth }\SpecialCharTok{+}\NormalTok{ depth }\SpecialCharTok{+}\NormalTok{ depth2 }\SpecialCharTok{+}\NormalTok{ distance\_to\_coast,}
\NormalTok{               D }\SpecialCharTok{\textasciitilde{}}\NormalTok{ depth }\SpecialCharTok{+}\NormalTok{ distance\_to\_coast,}
\NormalTok{               D }\SpecialCharTok{\textasciitilde{}}\NormalTok{ depth }\SpecialCharTok{+}\NormalTok{ distance\_to\_coast }\SpecialCharTok{+}\NormalTok{ distance\_to\_coast2,}
\NormalTok{               D }\SpecialCharTok{\textasciitilde{}}\NormalTok{ depth }\SpecialCharTok{+}\NormalTok{ depth2 }\SpecialCharTok{+}\NormalTok{ distance\_to\_coast,}
\NormalTok{               D }\SpecialCharTok{\textasciitilde{}}\NormalTok{ depth }\SpecialCharTok{+}\NormalTok{ depth2 }\SpecialCharTok{+}\NormalTok{ distance\_to\_coast }\SpecialCharTok{+}\NormalTok{ distance\_to\_coast2,}
\NormalTok{               D }\SpecialCharTok{\textasciitilde{}}\NormalTok{ depth }\SpecialCharTok{+}\NormalTok{ depth2 }\SpecialCharTok{+}\NormalTok{ distance\_to\_coast }\SpecialCharTok{+}\NormalTok{ distance\_to\_coast2 }\SpecialCharTok{+} 
\NormalTok{                 distance\_to\_coast3,}
\NormalTok{               D }\SpecialCharTok{\textasciitilde{}} \FunctionTok{s}\NormalTok{(depth, }\AttributeTok{k =} \DecValTok{3}\NormalTok{, }\AttributeTok{fx =} \ConstantTok{TRUE}\NormalTok{),}
\NormalTok{               D }\SpecialCharTok{\textasciitilde{}} \FunctionTok{s}\NormalTok{(depth, }\AttributeTok{k =} \DecValTok{4}\NormalTok{, }\AttributeTok{fx =} \ConstantTok{TRUE}\NormalTok{),}
\NormalTok{               D }\SpecialCharTok{\textasciitilde{}} \FunctionTok{s}\NormalTok{(depth, }\AttributeTok{k =} \DecValTok{5}\NormalTok{, }\AttributeTok{fx =} \ConstantTok{TRUE}\NormalTok{),}
\NormalTok{               D }\SpecialCharTok{\textasciitilde{}} \FunctionTok{s}\NormalTok{(depth, }\AttributeTok{k =} \DecValTok{6}\NormalTok{, }\AttributeTok{fx =} \ConstantTok{TRUE}\NormalTok{),}
\NormalTok{               D }\SpecialCharTok{\textasciitilde{}} \FunctionTok{s}\NormalTok{(depth, }\AttributeTok{k =} \DecValTok{7}\NormalTok{, }\AttributeTok{fx =} \ConstantTok{TRUE}\NormalTok{),}
\NormalTok{               D }\SpecialCharTok{\textasciitilde{}} \FunctionTok{s}\NormalTok{(depth, }\AttributeTok{k =} \DecValTok{8}\NormalTok{, }\AttributeTok{fx =} \ConstantTok{TRUE}\NormalTok{),}
\NormalTok{               D }\SpecialCharTok{\textasciitilde{}} \FunctionTok{s}\NormalTok{(depth, }\AttributeTok{k =} \DecValTok{6}\NormalTok{, }\AttributeTok{fx =} \ConstantTok{TRUE}\NormalTok{) }\SpecialCharTok{+}\NormalTok{ distance\_to\_coast,}
\NormalTok{               D }\SpecialCharTok{\textasciitilde{}} \FunctionTok{s}\NormalTok{(depth, }\AttributeTok{k =} \DecValTok{6}\NormalTok{, }\AttributeTok{fx =} \ConstantTok{TRUE}\NormalTok{) }\SpecialCharTok{+}\NormalTok{ distance\_to\_coast }\SpecialCharTok{+} 
\NormalTok{                 distance\_to\_coast2,}
\NormalTok{               D }\SpecialCharTok{\textasciitilde{}} \FunctionTok{s}\NormalTok{(distance\_to\_coast, }\AttributeTok{k =} \DecValTok{3}\NormalTok{, }\AttributeTok{fx =} \ConstantTok{TRUE}\NormalTok{),}
\NormalTok{               D }\SpecialCharTok{\textasciitilde{}} \FunctionTok{s}\NormalTok{(distance\_to\_coast, }\AttributeTok{k =} \DecValTok{4}\NormalTok{, }\AttributeTok{fx =} \ConstantTok{TRUE}\NormalTok{),}
\NormalTok{               D }\SpecialCharTok{\textasciitilde{}} \FunctionTok{s}\NormalTok{(distance\_to\_coast, }\AttributeTok{k =} \DecValTok{5}\NormalTok{, }\AttributeTok{fx =} \ConstantTok{TRUE}\NormalTok{),}
\NormalTok{               D }\SpecialCharTok{\textasciitilde{}} \FunctionTok{s}\NormalTok{(distance\_to\_coast, }\AttributeTok{k =} \DecValTok{6}\NormalTok{, }\AttributeTok{fx =} \ConstantTok{TRUE}\NormalTok{),}
\NormalTok{               D }\SpecialCharTok{\textasciitilde{}} \FunctionTok{s}\NormalTok{(distance\_to\_coast, }\AttributeTok{k =} \DecValTok{7}\NormalTok{, }\AttributeTok{fx =} \ConstantTok{TRUE}\NormalTok{),}
\NormalTok{               D }\SpecialCharTok{\textasciitilde{}} \FunctionTok{s}\NormalTok{(distance\_to\_coast, }\AttributeTok{k =} \DecValTok{8}\NormalTok{, }\AttributeTok{fx =} \ConstantTok{TRUE}\NormalTok{),}
\NormalTok{               D }\SpecialCharTok{\textasciitilde{}} \FunctionTok{s}\NormalTok{(distance\_to\_coast, }\AttributeTok{k =} \DecValTok{6}\NormalTok{, }\AttributeTok{fx =} \ConstantTok{TRUE}\NormalTok{) }\SpecialCharTok{+}\NormalTok{ depth,}
\NormalTok{               D }\SpecialCharTok{\textasciitilde{}} \FunctionTok{s}\NormalTok{(distance\_to\_coast, }\AttributeTok{k =} \DecValTok{6}\NormalTok{, }\AttributeTok{fx =} \ConstantTok{TRUE}\NormalTok{) }\SpecialCharTok{+}\NormalTok{ depth }\SpecialCharTok{+}\NormalTok{ depth2,}
\NormalTok{               D }\SpecialCharTok{\textasciitilde{}} \FunctionTok{s}\NormalTok{(depth, }\AttributeTok{k =} \DecValTok{4}\NormalTok{, }\AttributeTok{fx =} \ConstantTok{TRUE}\NormalTok{) }\SpecialCharTok{+} \FunctionTok{s}\NormalTok{(distance\_to\_coast, }\AttributeTok{k =} \DecValTok{4}\NormalTok{, }
                                                  \AttributeTok{fx =} \ConstantTok{TRUE}\NormalTok{),}
\NormalTok{               D }\SpecialCharTok{\textasciitilde{}} \FunctionTok{s}\NormalTok{(depth, }\AttributeTok{k =} \DecValTok{6}\NormalTok{, }\AttributeTok{fx =} \ConstantTok{TRUE}\NormalTok{) }\SpecialCharTok{+} \FunctionTok{s}\NormalTok{(distance\_to\_coast, }\AttributeTok{k =} \DecValTok{6}\NormalTok{, }
                                                  \AttributeTok{fx =} \ConstantTok{TRUE}\NormalTok{),}
\NormalTok{               D }\SpecialCharTok{\textasciitilde{}}\NormalTok{ distance\_to\_coast }\SpecialCharTok{+}\NormalTok{ depth }\SpecialCharTok{+}\NormalTok{ depth}\SpecialCharTok{:}\NormalTok{distance\_to\_coast,}
\NormalTok{               D }\SpecialCharTok{\textasciitilde{}}\NormalTok{ distance\_to\_coast }\SpecialCharTok{+}\NormalTok{ distance\_to\_coast2 }\SpecialCharTok{+}\NormalTok{ depth }\SpecialCharTok{+}\NormalTok{ depth2 }\SpecialCharTok{+} 
\NormalTok{                 depth}\SpecialCharTok{:}\NormalTok{distance\_to\_coast)}\ErrorTok{)}
\end{Highlighting}
\end{Shaded}

An overview of the model fits is presented below. We derive the AIC,
AICc and BIC for all models, and order based on AIC. AIC assumes
independence between the observation (as it is a likelihood based
statistic) and penalises for amount of parameters in the model. However,
if the log likelihood is overestimated (and thus the information) due to
unaccounted for dependence between the (some of) the variables, AIC
could incorrectly prefer more complex models. We acknowledge this,
however, we still use it as the case study was mainly functions to show
the functionality of the method -- if one would be explicitly interested
in the actual call abundance, finding a more effective model selection
criteria would be more important. In the following table, `ID' is a
unique identifier, `Density Model' is the \texttt{R} specification of
the density model, `N\_hat' is the estimated abundance, `\#par' is the
number of model parameters, and `AIC' is the AIC score \citep{Akaike1998Information}).

\begin{longtable}[]{@{}
  >{\raggedleft\arraybackslash}p{(\columnwidth - 8\tabcolsep) * \real{0.0261}}
  >{\raggedright\arraybackslash}p{(\columnwidth - 8\tabcolsep) * \real{0.7913}}
  >{\raggedleft\arraybackslash}p{(\columnwidth - 8\tabcolsep) * \real{0.0783}}
  >{\raggedleft\arraybackslash}p{(\columnwidth - 8\tabcolsep) * \real{0.0435}}
  >{\raggedleft\arraybackslash}p{(\columnwidth - 8\tabcolsep) * \real{0.0609}}@{}}
\toprule
\begin{minipage}[b]{\linewidth}\raggedleft
ID
\end{minipage} & \begin{minipage}[b]{\linewidth}\raggedright
Density Model
\end{minipage} & \begin{minipage}[b]{\linewidth}\raggedleft
N\_hat
\end{minipage} & \begin{minipage}[b]{\linewidth}\raggedleft
\#par
\end{minipage} & \begin{minipage}[b]{\linewidth}\raggedleft
AIC
\end{minipage} \\
\midrule
\endhead
33 & D \textasciitilde{} s(depth, k = 6, fx = TRUE) + s(distance\_to\_coast, k = 6,      fx = TRUE) & 5741.387 & 19 & 5894.555\\

30 & D \textasciitilde{} s(distance\_to\_coast, k = 6, fx = TRUE) + depth & 5141.567 & 15 & 5907.641\\

31 & D \textasciitilde{} s(distance\_to\_coast, k = 6, fx = TRUE) + depth + depth2 & 4029.957 & 16 & 5914.605\\

32 & D \textasciitilde{} s(depth, k = 4, fx = TRUE) + s(distance\_to\_coast, k = 4,      fx = TRUE) & 5974.553 & 15 & 5920.338\\

8 & D \textasciitilde{} logdepth + distance\_to\_coast + distance\_to\_coast2 + distance\_to\_coast3 & 7270.064 & 13 & 5944.458\\

28 & D \textasciitilde{} s(distance\_to\_coast, k = 7, fx = TRUE) & 6318.691 & 15 & 5948.720\\

29 & D \textasciitilde{} s(distance\_to\_coast, k = 8, fx = TRUE) & 6274.605 & 16 & 5950.358\\

15 & D \textasciitilde{} depth + depth2 + distance\_to\_coast + distance\_to\_coast2 +      distance\_to\_coast3 & 6263.857 & 14 & 5970.172\\

27 & D \textasciitilde{} s(distance\_to\_coast, k = 6, fx = TRUE) & 4977.064 & 14 & 5976.506\\

26 & D \textasciitilde{} s(distance\_to\_coast, k = 5, fx = TRUE) & 6112.981 & 13 & 5989.399\\

25 & D \textasciitilde{} s(distance\_to\_coast, k = 4, fx = TRUE) & 6647.313 & 12 & 5991.375\\

3 & D \textasciitilde{} distance\_to\_coast + distance\_to\_coast2 + distance\_to\_coast3 & 6012.470 & 12 & 6021.897\\

14 & D \textasciitilde{} depth + depth2 + distance\_to\_coast + distance\_to\_coast2 & 5053.378 & 13 & 6054.787\\

35 & D \textasciitilde{} distance\_to\_coast + distance\_to\_coast2 + depth + depth2 +      depth:distance\_to\_coast & 5117.734 & 14 & 6056.058\\

23 & D \textasciitilde{} s(depth, k = 6, fx = TRUE) + distance\_to\_coast + distance\_to\_coast2 & 5285.996 & 16 & 6059.020\\

9 & D \textasciitilde{} logdepth + depth + distance\_to\_coast + distance\_to\_coast2 & 4959.211 & 13 & 6059.691\\

12 & D \textasciitilde{} depth + distance\_to\_coast + distance\_to\_coast2 & 5185.959 & 12 & 6061.159\\

24 & D \textasciitilde{} s(distance\_to\_coast, k = 3, fx = TRUE) & 5912.876 & 11 & 6061.526\\

2 & D \textasciitilde{} distance\_to\_coast + distance\_to\_coast2 & 6050.358 & 11 & 6084.475\\

34 & D \textasciitilde{} distance\_to\_coast + depth + depth:distance\_to\_coast & 5024.660 & 12 & 6109.612\\

22 & D \textasciitilde{} s(depth, k = 6, fx = TRUE) + distance\_to\_coast & 6015.534 & 15 & 6173.505\\

10 & D \textasciitilde{} logdepth + depth + depth2 + distance\_to\_coast & 8430.175 & 13 & 6178.890\\

20 & D \textasciitilde{} s(depth, k = 7, fx = TRUE) & 8826.319 & 15 & 6246.687\\

19 & D \textasciitilde{} s(depth, k = 6, fx = TRUE) & 10760.817 & 14 & 6250.610\\

21 & D \textasciitilde{} s(depth, k = 8, fx = TRUE) & 9122.837 & 16 & 6252.560\\

13 & D \textasciitilde{} depth + depth2 + distance\_to\_coast & 11339.945 & 12 & 6273.316\\

7 & D \textasciitilde{} logdepth + depth + depth2 & 9691.473 & 12 & 6277.538\\

18 & D \textasciitilde{} s(depth, k = 5, fx = TRUE) & 23712.896 & 13 & 6305.295\\

5 & D \textasciitilde{} depth + depth2 & 113059.508 & 11 & 6325.998\\

17 & D \textasciitilde{} s(depth, k = 4, fx = TRUE) & 25480.347 & 12 & 6349.512\\

16 & D \textasciitilde{} s(depth, k = 3, fx = TRUE) & 30564.707 & 11 & 6362.469\\

6 & D \textasciitilde{} logdepth & 37011.682 & 10 & 6381.441\\

1 & D \textasciitilde{} 1 & 26638.358 & 9 & 6398.784\\

11 & D \textasciitilde{} depth + distance\_to\_coast & 27432.398 & 11 & 6400.660\\

4 & D \textasciitilde{} depth & 27365.887 & 10 & 6400.743\\
\bottomrule
\end{longtable}

\subsection{Results from the 999
bootstraps}\label{tab:results-from-the-999-bootstraps}

To find the uncertainty around our point estimates from Model 31, we
bootstrapped (i.e., re-sampled with replacement) the original calls 999
times. This creates 999 `equally likely' data sets, assuming that
calls are independent. The results from these bootstraps are presented
below. We present the parameter estimate, the lower and upper limits of
the absolute (LCL and UCL) and relative (LRCL and URCL) 95\% confidence
intervals, the standard deviation (SD) and the variance, and the
coefficient of variation (CV). Estimates are rounded to two significant
figures.

\begin{longtable}[]{@{}
  >{\raggedright\arraybackslash}p{(\columnwidth - 16\tabcolsep) * \real{0.2989}}
  >{\raggedright\arraybackslash}p{(\columnwidth - 16\tabcolsep) * \real{0.0805}}
  >{\raggedleft\arraybackslash}p{(\columnwidth - 16\tabcolsep) * \real{0.0805}}
  >{\raggedleft\arraybackslash}p{(\columnwidth - 16\tabcolsep) * \real{0.0805}}
  >{\raggedleft\arraybackslash}p{(\columnwidth - 16\tabcolsep) * \real{0.0920}}
  >{\raggedleft\arraybackslash}p{(\columnwidth - 16\tabcolsep) * \real{0.1034}}
  >{\raggedleft\arraybackslash}p{(\columnwidth - 16\tabcolsep) * \real{0.0920}}
  >{\raggedleft\arraybackslash}p{(\columnwidth - 16\tabcolsep) * \real{0.0805}}
  >{\raggedleft\arraybackslash}p{(\columnwidth - 16\tabcolsep) * \real{0.0920}}@{}}
\toprule
\begin{minipage}[b]{\linewidth}\raggedright
Par.
\end{minipage} & \begin{minipage}[b]{\linewidth}\raggedright
Link
\end{minipage} & \begin{minipage}[b]{\linewidth}\raggedleft
Est.
\end{minipage} & \begin{minipage}[b]{\linewidth}\raggedleft
LCL
\end{minipage} & \begin{minipage}[b]{\linewidth}\raggedleft
UCL
\end{minipage} & \begin{minipage}[b]{\linewidth}\raggedleft
LRCL
\end{minipage} & \begin{minipage}[b]{\linewidth}\raggedleft
URCL
\end{minipage} & \begin{minipage}[b]{\linewidth}\raggedleft
SD
\end{minipage} & \begin{minipage}[b]{\linewidth}\raggedleft
CV (\%)
\end{minipage} \\
\midrule
\endhead
$N$ & iden. & 5741 & 3877 & 5935 & -0.32 & 0.034 & 576 & 11\\
$g_0$ & logit & 0.20 & 0.072 & 0.36 & -0.65 & 0.74 & 0.061 & 29\\
$\beta_r$ & log & 2.8 & 2.8 & 2.9 & -0.017 & 0.017 & 0.019 & 0.67\\
$\sigma_r$ & log & 1.0 & 0.92 & 1.1 & -0.093 & 0.089 & 0.040 & 4.0\\
$\mu_s$ & log & 5.1 & 5.1 & 5.1 & -0.0030 & 0.0054 & 0.0094 & 0.19\\
$\sigma_s$ & log & 1.7 & 1.5 & 1.8 & -0.096 & 0.040 & 0.051 & 3.0\\
$\kappa $& log & -0.27 & -0.94 & 0.18 & -2.5 & 1.7 & 7.4 & 1423\\
$\delta_\kappa $& log & 3.8 & 3.7 & 4.0 & -0.028 & 0.044 & 0.064 & 1.7\\
$\psi_\kappa$ & logit & -2.2 & -2.4 & -1.9 & -0.12 & 0.12 & 0.11 & 4.9\\
$\beta_\text{(Intercept)}$ & log & -266 & -267 & -241 & -0.0062 & 0.092 & 7.9 & 3.0\\
$\beta_\text{s(depth).1}$ & log & 42 & 20 & 44 & -0.54 & 0.035 & 18 & 46\\
$\beta_\text{s(depth).2}$ & log & 302 & 293 & 304 & -0.032 & 0.0069 & 6.7 & 2.2 \\
$\beta_\text{s(depth).3}$ & log & 41 & -24 & 50 & -1.6 & 0.22 & 22 & 58 \\
$\beta_\text{s(depth).4}$ & log & -142 & -145 & -124 & -0.026 & 0.12 & 6.9 & 4.9 \\
$\beta_\text{s(depth).5}$ & log & 155 & 136 & 158 & -0.12 & 0.020 & 6.2 & 4.0 \\
$\beta_\text{s(distance\_to\_coast).1}$ & log & 159 & 145 & 162 & -0.089 & 0.016 & 4.5 & 2.8 \\
$\beta_\text{s(distance\_to\_coast).2}$ & log & -450 & -453 & -407 & -0.0083 & 0.095 & 16 & 3.5 \\
$\beta_\text{s(distance\_to\_coast).3}$ & log & 101 & 90 & 104 & -0.11 & 0.029 & 4.3 & 4.3 \\
$\beta_\text{s(distance\_to\_coast).4}$ & log & -109 & -110 & -97 & -0.011 & 0.12 & 5.1 & 4.7 \\
$\beta_\text{s(distance\_to\_coast).5}$ & log & -225 & -231 & -199 & -0.029 & 0.12 & 11 & 5.0 \\
\bottomrule
\end{longtable}

A surprising result is the high CV of \(\widehat{\log(\psi_\kappa)}\)
(1423\%). This is caused by a negative outlier (-234.24) and a mean close to 0. 
As these estimates are on the log scale and the outlier is \(<<0\), these effects are dampened when
converting to the real scale, which is why
\(\text{CV}(\widehat{\kappa}) = 22%
\) (see Table 2 in the main paper).

\newpage

\section{Appendix F: SNR
Likelihood}\label{appendix-f-snr-likelihood}

When noise is (highly) variable, truncating the data on the highest
noise level can result in dropping a substantial amount of data, which
can be a challenge when data are scarce. One solution, is to include
noise in the model and work with a signal-to-noise ratio (SNR) detection
function. Here we give details of such a model.\\
We first derive the full likelihood, for which we use several sources of
recorded information: the detection histories, the bearings, the
received levels and the noise levels. We assume that calls are
independent of each other in all their characteristics. For ease of
understanding, we first derive the likelihood for a single call. From
there, we expand it to include all calls.

We start with the joint distribution of our variables for a single call,
conditional on that the call involved at least two sensors, and on the
background noise. We assume that the detections, bearings and received
levels are all stochastic variables. The detection history is denoted by
the vector \(\boldsymbol{\omega}\), where element \(\omega_j = 1\) if
the call was detected at sensor \(j\), and \(0\) otherwise. We define
\(\omega^* = \sum_{j = 1}^K{\omega_j}\) , where \(K\) is the total
number of sensors in the array. Bearings are denoted by the vector
\(\boldsymbol{y}\), where element \(y_j\) is the bearing from sensor
\(j\) to the call in radians clockwise relative to north if the call was
detected, and \texttt{NA} otherwise. The received levels are denoted by
vector \(\boldsymbol{r}\), where element \(\boldsymbol{r}_j\) is the
root-mean-square (RMS) in dB re 1 \(\mu\)Pa of the received level of the
call at sensor \(j\) if the call was detected, and \texttt{NA}
otherwise. Noise is denoted by vector \(\boldsymbol{c}\), where element
\(c_j\) is the RMS in dB re 1 \(\mu\)Pa of the noise at sensor \(j\),
measured over the same frequency band as the call. The noise at sensors
that were involved with the detection was calculated over a 6-second
window consisting of the 3-seconds before and after the call; the noise
at sensors that were not involved was derived from a 6-second window
around the expected time of arrival of the call at that sensor.

\subsection{The detection probability for sensor \(j\),
\(p(r_j, c_j)\)}\label{the-detection-function-for-sensor-j-p_js-boldsymbolx-c}

Before we continue deriving the joint distribution for these stochastic
variables, we need to define the detection probability. Based on
preliminary research and knowledge of the sensors, we propose a
detection probability that is a function of signal-to-noise ratio, which is
the ratio between the received level and the noise. The noise showed
spatial and temporal variation and was assumed known without error.

Unlike the main paper, we do not propose a step function with a single detection probability, but a more
complex, gradually increasing detection probability. The simplicity of a
step function might be preferable when data are scarce and/or many
parameters are to be estimated, but a gradual function is likely to be
closer to the truth, as acoustics that interfere with detection could be
become loss relevance as SNR increases. Minimum requirements for that
detection probability are that it is monotonically increasing function of
SNR and bounded by 0 and 1. It should also allow for an upper asymptote
smaller than one, as we want to allow for non-perfect detection even at
(very) high values of SNR. Exploratory analysis showed that detection
probability could be well below 1 even for calls with high SNR. The
Janoschek function was used and denoted as follows: \begin{equation}
p( r_j, c_j ) = \theta_U - (\theta_U - \theta_L)\exp\left\{-\theta_R \times (r_j - c_j) ^{\theta_I}\right\},
\end{equation} where \(\theta_L\) and \(\theta_U\) are the respective
lower and upper asymptote, \(\theta_R > 0\) is the rate of increase,
\(\theta_I > 1\) controls the inflection point, and \(r_j - c_j\) is the
SNR, which is a subtraction as \(r\) and \(c\) are denoted on the
decibel scale. Setting the lower limit \(\theta_L\) to 0 this function
simplifies to \begin{equation}
p( r_j, c_j )  = \theta_U \left(1 -\exp\left\{-\theta_R \times  (r_j - c_j)^{\theta_I}\right\}\right). 
\end{equation} It is fundamental to understand that we often cannot use
the recordings to measure SNR or received level directly, as we only
have those for sensors that actually detect the call. There are \(K\)
sensors and generally only some detected the call. To evaluate the
likelihood of detection and of non-detection, we need a unique detection
probability for every sensor, which means we need an explanatory
variable that is different for every sensor. Thus, we use the
expectation of the SNR to determine the probability of detection at a
sensor, taking the other information as given (see Equation 24). 

As was already clear, we have two latent variables: the source level and
spatial origin of the call. We deal with this by marginalising over these
variables, as described below.

\subsection{Joint distribution of the detection
histories, bearings and received levels, given two or more detections
and noise,
\(f(\boldsymbol{\omega}, \boldsymbol{y}, \boldsymbol{r}| \omega^* \geq 2, \boldsymbol{c})\)}\label{joint-distribution-of-the-detection-histories-bearings-and-received-levels-given-two-or-more-detections-and-noise-fboldsymbolomega-boldsymboly-boldsymbolr-omega-geq-2-boldsymbolc}

We start by specifying the joint distribution of our observed stochastic
variables. We know this distribution does not only depend on detection,
but also on spatial origin and the source level of the detected call.
Since these variables are latent, we integrate them out, which gives the
following formula for the joint distribution \begin{equation}
f(\boldsymbol{\omega}, \boldsymbol{y}, \boldsymbol{r}| \omega^* \geq 2, \boldsymbol{c}) = \int_S\int_{\mathbb{R}^2}f(\boldsymbol{\omega}, \boldsymbol{y}, \boldsymbol{r}, \boldsymbol{x}, s| \omega^* \geq 2, \boldsymbol{c})d\boldsymbol{x}ds,
\end{equation} where \(\mathcal{S}\) is the support for \(s\), which in
theory is \((0, \infty )\) but in practice it suffices to use a smaller
support such as long as it sufficiently covers the probability density
of \(s\). Using standard conditional probability theory we can rewrite
this distribution as \begin{equation}
\begin{aligned}
\int_S\int_{\mathbb{R}^2}f(\boldsymbol{\omega}, \boldsymbol{y}, \boldsymbol{r}, \boldsymbol{x}, s| \omega^* \geq 2, \boldsymbol{c})d\boldsymbol{x}ds &= \\
\int_S\int_{\mathbb{R}^2}f(\boldsymbol{\omega}, \boldsymbol{y}, \boldsymbol{r}|\boldsymbol{x},& s, \omega^* \geq 2, \boldsymbol{c})f(\boldsymbol{x}| s, \omega^* \geq 2, \boldsymbol{c})  f(s|\omega^* \geq 2, \boldsymbol{c})d\boldsymbol{x}ds.
\end{aligned}
\end{equation} We will now focus on each component in Equation 6
separately, starting with
\(f(\boldsymbol{x}, s|\omega^* \geq 2, \boldsymbol{c}) = f(\boldsymbol{x}|s, \omega^* \geq 2, \boldsymbol{c})f(s|\omega^* \geq 2, \boldsymbol{c})\).

\subsection{Distribution of the source level of the
call, given at least two detections and noise,
\(f(s|\omega^* \geq 2, \boldsymbol{c})\)}\label{distribution-of-the-source-level-of-the-call-given-at-least-two-detections-and-noise-fsomega-geq-2-boldsymbolc}

Using Bayes' formula and the addition of latent variables similar to
before, we can rewrite the distribution of
\(s|\omega^* \geq 2, \boldsymbol{c}\) the following way \begin{equation}
f(s|\omega^* \geq 2, \boldsymbol{c}) = \frac{f(\omega^* \geq 2|s, \boldsymbol{c})f(s)}{f(\omega^* \geq 2| \boldsymbol{c})} = \frac{\int_{\mathbb{R}^2}f(\omega^* \geq 2|\boldsymbol{x}, s, \boldsymbol{c})f(\boldsymbol{x})d\boldsymbol{x}\times f(s)}{\int_S\int_{\mathbb{R}^2} f(\omega^* \geq 2| \boldsymbol{x}, s, \boldsymbol{c}) f(\boldsymbol{x}) f(s)d\boldsymbol{x}ds}.
\end{equation} Note that some probability distributions, such as the one
for source level, are not denoted as conditional on noise. This is
because the variables \(s\) and \(\boldsymbol{x}\) are assumed
independent of noise. (Actually, a recent paper by \cite{Thode2020Roaring}
showed that the source level of bowhead whale vocalisations in related
to other discrete sounds and noise, but here we assume them to be
independent.)

\subsection{Distribution of the spatial origin and
source level of the call, given at least two detections, noise, and
source level,
\(f(\boldsymbol{x}|\omega^* \geq 2, \boldsymbol{c}, s)\)}\label{distribution-of-the-spatial-origin-and-source-level-of-the-call-given-at-least-two-detections-noise-and-source-level-fboldsymbolxomega-geq-2-boldsymbolc-s}

The probability density of the spatial origin of an \emph{emitted} call
\(\boldsymbol{x}\) is proportional to the density at \(\boldsymbol{x}\),
such that \(f(\boldsymbol{x}) \propto D(\boldsymbol{x})\). The
normalising constant is therefore the integral over space,
\(\int_{\mathbb{R}^2} D(\boldsymbol{x}) d\boldsymbol{x}\) and this gives
\begin{equation}
f(\boldsymbol{x}) = \frac{D(\boldsymbol{x})}{\int_{\mathbb{R}^2} D(\boldsymbol{x}) d\boldsymbol{x}}.
\end{equation} We have not defined a spatial density model
\(D(\boldsymbol{x})\) for the calls yet. It is common practice to try
several density models and then choose to best one using some criterion
such as AIC. The density of
\(\boldsymbol{x}| s, \omega^* \geq 2, \boldsymbol{c}\) is
\begin{equation}
f(\boldsymbol{x}| s, \omega^* \geq 2, \boldsymbol{c})  = \frac{f(\omega^* \geq 2| \boldsymbol{x}, s, \boldsymbol{c})f(\boldsymbol{x}| s, \boldsymbol{c})}{f(\omega^* \geq 2| s, \boldsymbol{c})} = \frac{f(\omega^* \geq 2| \boldsymbol{x}, s, \boldsymbol{c})f(\boldsymbol{x})}{\int_{\mathbb{R}^2}f(\omega^* \geq 2|\boldsymbol{x}, s, \boldsymbol{c})f(\boldsymbol{x})d\boldsymbol{x}}.
\end{equation} In both definitions we assume the distribution of
\emph{emitted} calls to be unrelated to \(s\) and \(\boldsymbol{c}\),
thus \(f(\boldsymbol{x}|s, \boldsymbol{c}) \equiv f(\boldsymbol{x})\)
(note that this is not true for the distribution of \emph{detected}
calls, which changes when \(s\) and \(\boldsymbol{c}\) changes). As
noted before, in this study, we assume
\(s \sim \mathcal{N}(\mu_s, \sigma_s^2)\).

Combining Equations 8 and 9 gives \begin{equation}
\begin{aligned}
f(\boldsymbol{x}| s, \omega^* \geq 2,  & \boldsymbol{c}) \times f(s|\omega^* \geq 2, \boldsymbol{c})\\ 
& = \frac{f(\omega^* \geq 2| \boldsymbol{x}, s, \boldsymbol{c})f(\boldsymbol{x})}{\int_{\mathbb{R}^2}f(\omega^* \geq 2|\boldsymbol{x}, s, \boldsymbol{c})f(\boldsymbol{x})d\boldsymbol{x}} \times \frac{\int_{\mathbb{R}^2}f(\omega^* \geq 2|\boldsymbol{x}, s, \boldsymbol{c})f(\boldsymbol{x})d\boldsymbol{x}f(s)}{\int_S\int_{\mathbb{R}^2} f(\omega^* \geq 2| \boldsymbol{x}, s, \boldsymbol{c}) f(\boldsymbol{x}) f(s)d\boldsymbol{x}ds} \\ 
& = \frac{f(\omega^* \geq 2| \boldsymbol{x}, s, \boldsymbol{c})f(s)}{\int_S\int_{\mathbb{R}^2} f(\omega^* \geq 2| \boldsymbol{x}, s, \boldsymbol{c}) \frac{D(\boldsymbol{x})}{\int_{\mathbb{R}^2} D(\boldsymbol{x}) d\boldsymbol{x}} f(s)d\boldsymbol{x}ds} \times \frac{D(\boldsymbol{x})}{\int_{\mathbb{R}^2} D(\boldsymbol{x}) d\boldsymbol{x}} \\ 
& = \frac{f(\omega^* \geq 2| \boldsymbol{x}, s, \boldsymbol{c})D(\boldsymbol{x})f(s)}{a},
\end{aligned}
\end{equation} where in the last step the constants
\({\int_{\mathbb{R}^2} D(\boldsymbol{x}) d\boldsymbol{x}}\) cancel out
and
\(a \coloneqq \int_S\int_{\mathbb{R}^2} f(\omega^* \geq 2| \boldsymbol{x}, s, \boldsymbol{c}) D(\boldsymbol{x}) f(s)d\boldsymbol{x}ds\).

Finally, the distribution of
\(f(\omega^* \geq 2| \boldsymbol{x}, s, \boldsymbol{c})\) is a
probability mass function, where \(\omega^* \geq 2\) can either attain
the value \(0\) when detected at most once or \(1\) when detected twice
or more. This is represented by the following: \begin{equation}
\begin{aligned}
f(\omega^* \geq 2| &\boldsymbol{x}, s, \boldsymbol{c}) \\ 
& =  1 - \mathbb{P}(\omega^* = 0| \boldsymbol{x}, s, \boldsymbol{c}) - \mathbb{P}(\omega^*= 1| \boldsymbol{x}, s, \boldsymbol{c})\\
& = 1 - \prod_{j=1}^K \left(1-g_j(\bx, s, c_j) \right) - \sum^K_{j=1} \left( g_j(\bx, s, c_j) \prod_{k \in \{1:K|k \neq j\}} \left(1-g_k(\bx, s, c_k) \right) \right),
\end{aligned}
\end{equation}
where $g_j(\bx, s, c_j)$ denotes the detection function at detector $j$, which is specified in Equation 24.
Equation 9 can readily be extended to set the minimum number of
detections per call at 3, 4, etc. (although the formula and
computational complexity significantly increases; see the Discussion in
the main paper).

\subsection{Joint distribution of the capture
history, bearings and received level for a call, given its spatial
origin, source level, at least two detections and noise,
\(f(\boldsymbol{\omega}, \boldsymbol{y}, \boldsymbol{r}| \boldsymbol{x}, s, \omega^* \geq 2, \boldsymbol{c})\)}\label{joint-distribution-of-the-capture-history-bearings-and-received-level-for-a-call-given-its-spatial-origin-source-level-at-least-two-detections-and-noise-fboldsymbolomega-boldsymboly-boldsymbolr-boldsymbolx-s-omega-geq-2-boldsymbolc}

First, we observe that \begin{equation}
\begin{aligned}
& f(\boldsymbol{\omega}, \boldsymbol{y}, \boldsymbol{r}| \boldsymbol{x}, s, \omega^* \geq 2, \boldsymbol{c}) \\ &\quad\quad =  f(\boldsymbol{\omega}| \boldsymbol{x}, s, \omega^* \geq 2, \boldsymbol{c}) \times f(\boldsymbol{y}| \boldsymbol{\omega}, \boldsymbol{x}, s, \omega^* \geq 2, \boldsymbol{c}) \times f(\boldsymbol{r}| \boldsymbol{y}, \boldsymbol{\omega}, \boldsymbol{x}, s, \omega^* \geq 2, \boldsymbol{c}).
\end{aligned}
\end{equation} Since \(\boldsymbol{y}\) and \(s\) are unrelated, and so
are \(\boldsymbol{r}\) and \(\boldsymbol{y}\), we can rewrite the above
formula as \begin{equation}
\begin{aligned}
& f(\boldsymbol{\omega}, \boldsymbol{y}, \boldsymbol{r}| \boldsymbol{x}, s, \omega^* \geq 2, \boldsymbol{c}) \\ &\quad\quad = f(\boldsymbol{\omega}| \boldsymbol{x}, s, \omega^* \geq 2, \boldsymbol{c}) \times f(\boldsymbol{y}| \boldsymbol{\omega}, \boldsymbol{x}, \omega^* \geq 2, \boldsymbol{c}) \times f(\boldsymbol{r}| \boldsymbol{\omega}, \boldsymbol{x}, s, \omega^* \geq 2, \boldsymbol{c}).
\end{aligned}
\end{equation}

We now evaluate this joint density from left to right, starting with the
density of
\(\boldsymbol{\omega}| \boldsymbol{x}, s, \omega^* \geq 2, \boldsymbol{c}\).
First, we rewrite
\(f(\boldsymbol{\omega}| \boldsymbol{x}, s, \omega^* \geq 2, \boldsymbol{c})\)
as \begin{equation}
\begin{aligned}
f(\boldsymbol{\omega}| \boldsymbol{x}, s, \omega^* \geq 2, \boldsymbol{c}) &= \frac{f(\omega^* \geq 2| \boldsymbol{\omega}, \boldsymbol{x}, s, \boldsymbol{c}) f(\boldsymbol{\omega}| \boldsymbol{x}, s, \boldsymbol{c})}{f(\omega^* \geq 2| \boldsymbol{x}, s, \boldsymbol{c})} \\ &= \frac{1 \times f(\boldsymbol{\omega}| \boldsymbol{x}, s, \boldsymbol{c})}{f(\omega^* \geq 2| \boldsymbol{x}, s, \boldsymbol{c})} = \frac{f(\boldsymbol{\omega}| \boldsymbol{x}, s, \boldsymbol{c})}{f(\omega^* \geq 2| \boldsymbol{x}, s, \boldsymbol{c})}.
\end{aligned}
\end{equation} Conditioning on the detection history gives a probability
of two or more detections, since the probability of detection at least
twice is implied by the existence of the detection history.

We can view the detection history of a call as a set of Bernoulli trials
with size \(K\) and non-constant probability
\(p_j(\boldsymbol{x}, s, \boldsymbol{c})\) with \(j = 1,..,K\). This
gives \begin{equation}
f(\boldsymbol{\omega}| \boldsymbol{x}, s, \omega^* \geq 2, \boldsymbol{c}) =  \frac{\prod^K_{j = 1} g_j(\bx, s, c_j) ^ {\omega_j} (1 - g_j(\bx, s, c_j) ) ^ {1 - \omega_j}}{f(\omega^* \geq 2| \boldsymbol{x}, s, \boldsymbol{c})}.
\end{equation} The second term in Equation 14 is the density of
\(\boldsymbol{y}| \boldsymbol{\omega}, \boldsymbol{x}, \omega^* \geq 2, \boldsymbol{c}\),
which we can rewrite as \begin{equation}
\begin{aligned}
f(\boldsymbol{y}| \boldsymbol{\omega}, \boldsymbol{x}, \omega^* \geq 2, \boldsymbol{c}) &= \frac{f(\omega^* \geq 2| \boldsymbol{y}, \boldsymbol{\omega}, \boldsymbol{x}, \boldsymbol{c}) f(\boldsymbol{y}| \boldsymbol{\omega}, \boldsymbol{x})} {f(\omega^* \geq 2| \boldsymbol{\omega}, \boldsymbol{x}, \boldsymbol{c})}\\ &= \frac{1 \times f(\boldsymbol{y}| \boldsymbol{\omega}, \boldsymbol{x})} {1} =  f(\boldsymbol{y}| \boldsymbol{\omega}, \boldsymbol{x}),
\end{aligned}
\end{equation} where we assume independence between bearings and noise.

To keep this section as simple as possible, we assume that all bearings
follow a Von Mises distribution with a single concentration parameter
\(\kappa\) (unlike the main paper, where we assume a 2-part mixture),
which gives \begin{equation}
f(\boldsymbol{y}| \boldsymbol{\omega}, \boldsymbol{x}) = \prod_{j \in\{ 1:K| \omega_j = 1 \}} \left( \frac{\exp\{ \kappa  \cos(y_j - \mathop{\mathrm{\mathbb{E}}}[y_j | \boldsymbol{x}]) \}}{2 \pi I_0(\kappa)} \right),
\end{equation} where \(I_0\) is the modified Bessel function of order
\(0\) and \(\mathop{\mathrm{\mathbb{E}}}[y_j | \boldsymbol{x}]\) is the
expected bearing at sensor \(j\) for origin \(\boldsymbol{x}\).

Finally, we rewrite the density of
\(\boldsymbol{r}| \boldsymbol{\omega}, \boldsymbol{x}, s, \omega^* \geq 2, \boldsymbol{c}\)
as \begin{equation}
\begin{aligned}
f(\boldsymbol{r}| \boldsymbol{\omega}, \boldsymbol{x}, s, \omega^* \geq 2, \boldsymbol{c}) &= \frac{f(\omega^* \geq 2| \boldsymbol{r}, \boldsymbol{\omega}, \boldsymbol{x}, s, \boldsymbol{c})f(\boldsymbol{r}| \boldsymbol{\omega}, \boldsymbol{x}, s, \boldsymbol{c})} {f(\omega^* \geq 2| \boldsymbol{\omega}, \boldsymbol{x}, s, \boldsymbol{c})}\\ &= \frac{1 \times f(\boldsymbol{r}| \boldsymbol{\omega}, \boldsymbol{x}, s)} {1} = f(\boldsymbol{r}| \boldsymbol{\omega}, \boldsymbol{x}, s, \boldsymbol{c})
\end{aligned}
\end{equation} Following the model described in the main paper, we
assume a Gaussian error on received levels to capture measurement error
and error in the propagation model, such that \begin{equation}
r_j | \boldsymbol{x},s \sim \mathcal{N}(\mathop{\mathrm{\mathbb{E}}}[r_j | \boldsymbol{x},s], \sigma_r^2). 
\end{equation} Received level itself is a function of: source level,
denoted by \(s\); the Euclidean distance in meters, denoted by
\(d_j(\boldsymbol{x})\), between the spatial origin of the call
\(\boldsymbol{x}\) and sensor \(j\); and the transmission loss per
distance travelled by the sound, \(\beta_r\). Standard sound propagation
theory \citep{Jensen2011Computational} gives the following formula for the
expectation for \(r_j\): \begin{equation}
\mathop{\mathrm{\mathbb{E}}}[r_j|s, \boldsymbol{x}] = s - \beta_r\log_{10}\left( d_j(\boldsymbol{x}) \right),
\end{equation} where \(\beta_r\) is a parameter indexing the rate at
which received level decreases with distance. More complex propagation
models can be used, as was shown by \cite{Phillips2016Passive} and \cite{Royle2018Modelling}. As
\(r_j\) and \(c_j\) are both denoted on the dB scale, the expected
signal-to-noise ratio at sensor \(j\) is simply given by
\begin{equation}
\mathop{\mathrm{\mathbb{E}}}[\text{SNR}_j| s, \boldsymbol{x}, c_j ]=  \mathop{\mathrm{\mathbb{E}}}[r_j|s, \boldsymbol{x}] - c_j = s - \beta_r\log_{10}\left( d_j(\boldsymbol{x}) \right) - c_j.
\end{equation}

Signal strength is only recorded if the call is detected, and thus the
density is only specified when \(\omega_j = 1\). Assuming independence
between the detectors, we can write \begin{equation}
f(\boldsymbol{r}| \boldsymbol{\omega}, \boldsymbol{x}, s, \boldsymbol{c}) = \prod_{j \in\{ 1:K| \omega_j = 1 \}} f(r_j | \omega_j = 1,  \boldsymbol{x}, s, c_j),
\end{equation} where Bayes' formula gives \begin{equation}
f(r_j | \omega_j = 1,  \boldsymbol{x}, s, c_j) = \frac{f(\omega_j = 1 | r_j ,  \boldsymbol{x}, s, c_j) \times f(r_j | \boldsymbol{x}, s, c_j)}{f(\omega_j = 1 |\boldsymbol{x}, s, c_j)}
\end{equation} The first element of the numerator in Equation 21 is the
probability of detection given the received level and the noise, which
is given by Equation 4 but now with \(r_j\) known. This means that
\(\boldsymbol{x}\) and \(s\) are redundant, thus giving \begin{equation}
f(\omega_j = 1 | r_j ,  \boldsymbol{x}, s, c_j) = f(\omega_j = 1 | r_j , c_j) = p(r_j - c_j)
\end{equation} The second element of the numerator in Equation 21 is the
probability density function of received level given the spatial origin
and the source level, which is specified in Equation 19. As true
received level and noise are assumed independent, we get
\begin{equation}
f(r_j | \boldsymbol{x}, s, c_j) = f(r_j | \boldsymbol{x}, s) = \frac{1}{\sigma_r} \varphi\left(\frac{r_j - \mathop{\mathrm{\mathbb{E}}}(r_j|\boldsymbol{x}, s)}{\sigma_r} \right),
\end{equation} where \(\varphi\) denotes the standard normal probability
density function. The denominator in Equation 21 is the probability of
detection given the spatial origin, source level, and noise at sensor
\(j\). As we also need the received level to derive this probability, we
condition on and marginalise over \(r_j\), which gives \begin{equation}
\begin{aligned}
f(\omega_j = 1 |\boldsymbol{x}, s, c_j) &= \int_\mathcal{R} f(\omega_j = 1 |r, \boldsymbol{x}, s, c_j) f(r | \boldsymbol{x}, s, c_j)dr \\
&=  \int_\mathcal{R} p(r - c_j) \times \frac{1}{\sigma_r} \varphi\left(\frac{r - \mathop{\mathrm{\mathbb{E}}}(r|\boldsymbol{x}, s)}{\sigma_r} \right)dr,
\end{aligned}
\end{equation} where $\mathcal{R} \in (-\infty, \infty)$ denotes the support for the received level. 
Equation 24 is called the detection function and, in line with other literature, is also denoted by $g_j(\bx, s, c_j)$. 
We combine the three elements to rewrite Equation 20 to \begin{equation}
f(\boldsymbol{r}| \boldsymbol{\omega}, \boldsymbol{x}, s, \boldsymbol{c}) = \prod_{j \in\{ 1:K| \omega_j = 1 \}}
\frac{1}{\sigma_r} \times \frac{p(r_j - c_j) \times  \varphi\left(\frac{r_j - \mathop{\mathrm{\mathbb{E}}}(r_j|\boldsymbol{x}, s)}{\sigma_r} \right)}
{g_j(\bx, s, c_j)}
\end{equation}

Putting all of this together gives the following joint distribution
\begin{equation}
\begin{aligned}
&f(\boldsymbol{\omega}, \boldsymbol{y}, \boldsymbol{r}, | \omega^* \geq 2, \boldsymbol{c}) = \int_S\int_{\mathbb{R}^2} \frac{f(\omega^* \geq 2| \boldsymbol{x}, s, \boldsymbol{c})D(\boldsymbol{x})f(s)}{a} 
\\ & \quad\quad \times 
\frac{\prod^K_{j = 1} g_j(\bx, s, c_j) ^ {\omega_j} (1 - g_j(\bx, s, c_j)) ^ {1 - \omega_j}}{f(\omega^* \geq 2| \boldsymbol{x}, s, \boldsymbol{c})} \\ 
& \quad\quad \times 
\prod_{j \in\{ 1:K| \omega_j = 1 \}} \frac{\exp\{ \kappa \cos(y_j - \mathop{\mathrm{\mathbb{E}}}[y_j | \boldsymbol{x}]) \}}{2 \pi I_0(\kappa)}  \\
& \quad\quad \times
\prod_{j \in\{ 1:K| \omega_j = 1 \}}
 \frac{p(r_j - c_j) \times  \varphi\left(\frac{r_j - \mathop{\mathrm{\mathbb{E}}}(r_j|\boldsymbol{x}, s)}{\sigma_r} \right)}
{\int_\mathcal{R} p(r - c_j) \times \varphi\left(\frac{r - \mathop{\mathrm{\mathbb{E}}}(r|\boldsymbol{x}, s)}{\sigma_r} \right)dr} d\boldsymbol{x}ds \\ 
& \quad = \frac{1}{a} \int_S\int_{\mathbb{R}^2} D(\boldsymbol{x})f(s)\prod^K_{j = 1} g_j(\bx, s, c_j) ^ {\omega_j} (1 - g_j(\bx, s, c_j)) ^ {1 - \omega_j} \\
&\quad\quad \times \prod_{j \in\{ 1:K| \omega_j = 1 \}} 
 \frac{\exp\{ \kappa \cos(y_{j} - \mathop{\mathrm{\mathbb{E}}}[y_{j} | \boldsymbol{x}]) \} }{2 \pi I_0(\kappa)} 
\times
 \frac{p(r_j - c_j) \times  \varphi\left(\frac{r_j - \mathop{\mathrm{\mathbb{E}}}(r_j|\boldsymbol{x}, s)}{\sigma_r} \right)}
{\int_\mathcal{R} p(r - c_j) \times \varphi\left(\frac{r - \mathop{\mathrm{\mathbb{E}}}(r|\boldsymbol{x}, s)}{\sigma_r} \right)dr}    d\boldsymbol{x}ds.
\end{aligned}
\end{equation}

\subsection{Generalising to \(n\)
calls}\label{generalising-to-n-calls}

The above gave the density of the detection history, bearing and
received level for a single detection. Here we generalise to\(n\) calls
that were detected at least twice. We assume that all calls are
independent of one another (this assumption is certainly violated due to
spatial correlation in the individuals creating the calls). We define
the \((n\times2)\)-matrix of all spatial origins
\(\boldsymbol{X}= (\boldsymbol{x}_1,...,\boldsymbol{x}_n)\), the
\((n\times1)\)-vector of all source levels
\(\boldsymbol{s}= (s_1,...,s_n)\), the \((n\times K)\)-matrix of all
detection histories
\(\boldsymbol{\Omega}= (\boldsymbol{\omega}_1,...,\boldsymbol{\omega}_n)\),
the \((n\times K)\)-matrix of all bearings
\(\boldsymbol{Y}= (\boldsymbol{y}_1,...,\boldsymbol{y}_n)\), the
\((n\times K)\)-matrix of all received levels as
\(\boldsymbol{R}= (\boldsymbol{r}_1,...,\boldsymbol{r}_n)\) and the
\((n\times K)\)-matrix of all the noise as
\(\boldsymbol{C}= (\boldsymbol{c}_1,...,\boldsymbol{c}_n)\). Using this
notation, we can create the joint distribution of
\(n, \boldsymbol{\Omega},\boldsymbol{Y},\boldsymbol{R}| \boldsymbol{C}\)
as

\begin{equation}
\begin{aligned}
f(n, \boldsymbol{\Omega}, \boldsymbol{Y}, \boldsymbol{R}| \boldsymbol{C}) &= f(n| \boldsymbol{C}) f(\boldsymbol{\Omega}, \boldsymbol{Y}, \boldsymbol{R}| n, \boldsymbol{C}) \\
&= f(n )  \int_S\int_{\mathbb{R}^2} f(\boldsymbol{\Omega}, \boldsymbol{Y}, \boldsymbol{R}, \boldsymbol{X}, \boldsymbol{s}| n, \boldsymbol{C}) d\boldsymbol{X}d\boldsymbol{s}
\end{aligned}
\end{equation}

We show below why \(f(n|\boldsymbol{C}) \equiv f(n)\).

\subsection{Number of detected animals,
\(n\)}\label{number-of-detected-animals-n}

Originally in SCR, we assume that the calls occur in space according to
an inhomogeneous Poisson point process with intensity
\(D(\boldsymbol{x})\). As not all calls are detected, the
\emph{detected} calls occur in space according to a filtered
inhomogeneous Poisson point process with a rate parameter being a
function of \(\boldsymbol{x}\), i.e.~\(\lambda(\boldsymbol{x})\),
multiplied by the probability that this call was detected at all,
\(f(\omega^* \geq 2| \boldsymbol{x})\). The total number of detected
calls then follows a Poisson distribution with rate parameter equal to
the integral over (a subset of) \(\mathbb{R}^2\), giving
\(\lambda = \int_{\mathbb{R}^2} D(\boldsymbol{x}) f(\omega^* \geq 2| \boldsymbol{x}) d\boldsymbol{x}\).

We now added the random variable source level \(s\) to better explain
detection probability. Since the source level is latent, we assume a
distribution and integrate it out. This means that our detected calls
now follow an inhomogeneous Poisson point process with rate parameter
\(D(\boldsymbol{x}) = \int_S D(x | s) f(s) ds\), and our total number of
detected calls has a rate parameter
\(\lambda = \int_{\mathbb{R}^2} \int_S D(\boldsymbol{x}) f(s) f(\omega^* \geq 2| \boldsymbol{x}, s)ds d\boldsymbol{x}\).

So far, our spatial process was \emph{space-inhomogeneous}, but
\emph{time-homogeneous}. In fact, this is one of the assumptions of
Poisson processes --- their intensity or rate parameter is constant
throughout time. The information on noise, however, varies over time,
which means that our intensity parameter \(\lambda\) varies over time.
When a Poisson count process is time-inhomogeneous, the rate parameter
of Poisson process over the entire study period can be derived by
integrating over time. As our time-related variable is noise, this means
we have to integrate over noise. Assuming that we have a (systematic)
random sample of \(b\) noise recordings for all \(K\) sensors, we denote
the matrix of these noise recordings by \(\mathcal{C}\). The rate
parameter for the underlying Poisson point process thus is approximated
by,
\(\lambda \approx \frac{ \sum_{\boldsymbol{c}\in \mathcal{C}} \int_S D(\boldsymbol{x}) f(s) f(\omega^* \geq 2| \boldsymbol{x}, s ,\boldsymbol{c})ds }{b}\),
involving a Monte-Carlo integration over noise. The distribution of
\(n\) follows as \begin{equation}
f(n) \sim \text{Poisson}\left(\frac{ \int_{\mathbb{R}^2} \sum_{\boldsymbol{c}\in \mathcal{C}} \int_S D(\boldsymbol{x}) f(s) f(\omega^* \geq 2| \boldsymbol{x}, s ,\boldsymbol{c})ds d\boldsymbol{x}}{b} \right).
\end{equation}

\subsection{Joint distribution of all detection
histories, bearings, received levels, spatial origins and source levels,
given the detected calls,
\(\boldsymbol{\Omega}, \boldsymbol{Y}, \boldsymbol{R}, \boldsymbol{X}, \boldsymbol{s}| n, \boldsymbol{C}\)}\label{joint-distribution-of-all-detection-histories-bearings-received-levels-spatial-origins-and-source-levels-given-the-detected-calls-boldsymbolomega-boldsymboly-boldsymbolr-boldsymbolx-boldsymbols-n-boldsymbolc}

As we assume calls to be independent of each other, the part inside the
integrals becomes the product of the distribution for the individual
calls. Moreover, conditioning on \(n\) is equivalent to conditioning on
every individual call being detected at least twice (see \cite{Borchers2008Spatially}). This gives: \begin{equation}
\begin{aligned}
f(\boldsymbol{\Omega}, \boldsymbol{Y}, \boldsymbol{R}, &\boldsymbol{X}, \boldsymbol{s}| n, \boldsymbol{C}) \\
&= f(\boldsymbol{\Omega}| n, \boldsymbol{C}) f(\boldsymbol{Y}, \boldsymbol{R}, \boldsymbol{X}, \boldsymbol{s}| \boldsymbol{\Omega}, n, \boldsymbol{C}) \\ 
& \equiv \binom{n}{n_1,...,n_\mathcal{U}} \prod_{i=1}^n f(\boldsymbol{\omega}_i | \omega^* \geq 2_i,\boldsymbol{c}_i) f( \boldsymbol{y}_i, \boldsymbol{r}_i, \boldsymbol{x}_i, s_i| \boldsymbol{\omega}_i, \omega^* \geq 2_i,\boldsymbol{c}_i) \\
& =  \binom{n}{n_1,...,n_\mathcal{U}} \prod_{i=1}^n f(\boldsymbol{\omega}_i , \boldsymbol{y}_i, \boldsymbol{r}_i, \boldsymbol{x}_i, s_i| \omega^* \geq 2_i,\boldsymbol{c}_i).
\end{aligned}
\end{equation} with \(\binom{n}{n_1,...,n_\mathcal{U}}\) being the
multinomial coefficient where \(n_1,...,n_\mathcal{U}\) are the
frequencies for \(\mathcal{U}\) unique detection histories. This
coefficient appears as the order of the calls does not matter. Adding
the integrals again and using the independence assumption between calls,
we get the following \begin{equation}
\begin{aligned}
f(n) 
\times
\int_S&\int_{\mathbb{R}^2} f(\boldsymbol{\Omega}, \boldsymbol{Y}, \boldsymbol{R}, \boldsymbol{X}, \boldsymbol{s}| n, \boldsymbol{C}) d\boldsymbol{X}d\boldsymbol{s}\\
& = f(n)
{\binom{n}{n_1,...,n_\mathcal{U}}} \\
& \quad \times
\int_S \cdots \int_S \int_{\mathbb{R}^2} \cdots \int_{\mathbb{R}^2} \prod_{i = 1}^n f(\boldsymbol{\omega}_i, \boldsymbol{y}_i, \boldsymbol{r}_i, \boldsymbol{x}_i, s_i| \omega^* \geq 2_i,\boldsymbol{c}_i) d\boldsymbol{x}_1 \cdots  d\boldsymbol{x}_n ds_1 \cdots ds_n \\ 
& = f(n) 
{\binom{n}{n_1,...,n_\mathcal{U}}} \\
& \quad \times
\int_S\int_{\mathbb{R}^2} \cdots \int_S\int_{\mathbb{R}^2} \prod_{i = 1}^n f(\boldsymbol{\omega}_i, \boldsymbol{y}_i, \boldsymbol{r}_i, \boldsymbol{x}_i, s_i| \omega^* \geq 2_i,\boldsymbol{c}_i) d\boldsymbol{x}_1 ds_1 \cdots d\boldsymbol{x}_n ds_n \\ 
& = f(n) 
{\binom{n}{n_1,...,n_\mathcal{U}}} 
\prod_{i = 1}^n \int_S\int_{\mathbb{R}^2}   f(\boldsymbol{\omega}_i, \boldsymbol{y}_i, \boldsymbol{r}_i, \boldsymbol{x}_i, s_i|\omega^* \geq 2_i,\boldsymbol{c}_i) d\boldsymbol{x}_i ds_i,
\end{aligned}
\end{equation} where \begin{equation}
f(n) = \frac{ \lambda^n \exp\{ -\lambda \} }{n!}, \lambda = \frac{ \int_{\mathbb{R}^2} \sum_{\boldsymbol{c}\in \mathcal{C}} \int_S D(\boldsymbol{x}) f(s) f(\omega^* \geq 2| \boldsymbol{x}, s ,\boldsymbol{c})ds d\boldsymbol{x}}{b}.
\end{equation}

\subsection{Density of data, and full likelihood}

We now use the result from the previous section to obtain the density of
the data conditional on noise levels,
\(f(n, \boldsymbol{\Omega}, \boldsymbol{Y}, \boldsymbol{R}| \boldsymbol{C})\),
and hence obtain the full likelihood. \begin{equation}
\begin{aligned}
f(n, \boldsymbol{\Omega},& \boldsymbol{Y}, \boldsymbol{R}| \boldsymbol{C}) \\
& =  \int_S \int_{\mathbb{R}^2} f(n, \boldsymbol{\Omega}, \boldsymbol{Y}, \boldsymbol{R}, \boldsymbol{X}, \boldsymbol{s}| \boldsymbol{C}) d\boldsymbol{X}d\boldsymbol{s}\\
& = f(n)  \int_S \int_{\mathbb{R}^2} f(\boldsymbol{\Omega}, \boldsymbol{Y}, \boldsymbol{R}, \boldsymbol{X}, \boldsymbol{s}| n, \boldsymbol{C}) d\boldsymbol{X}d\boldsymbol{s}\\
& = \frac{ \lambda^n \exp\{ -\lambda \} }{n!}
{\binom{n}{n_1,...,n_\mathcal{U}}}
\prod_{i = 1}^n
\frac{1}{a_i} \int_S\int_{\mathbb{R}^2} D(\boldsymbol{x}_i)f(s_i)\prod^K_{j = 1} p_j(\boldsymbol{x}_i, s_i, c_{ij}) ^ {\omega_{ij}} (1 - p_j(\boldsymbol{x}_i, s_i, c_{ij})) ^ {1 - \omega_{ij}} \\
&\quad\quad \times \prod_{j \in\{ 1:K| \omega_j = 1 \}}
\left( \frac{\exp\{ \kappa \cos(y_{ij} - \mathop{\mathrm{\mathbb{E}}}[y_{ij} | \boldsymbol{x}_i]) \} }{2 \pi I_0(\kappa)}
\times
\frac{\varphi \left(\frac{r_{ij} - \mathop{\mathrm{\mathbb{E}}}[r_{ij}|\boldsymbol{x}_i, s_i, c_{ij}] }{\sigma_r}\right)}{ \sigma_r \left( 1 - \Phi \left(\frac{c_{ij} -  \mathop{\mathrm{\mathbb{E}}}[r_{ij}|\boldsymbol{x}_i, s_i, c_{ij}]}{\sigma_r}\right)\right)}   \right)  d\boldsymbol{x}_i ds_i,
\end{aligned}
\end{equation} where
\(a_i \coloneqq \int_S\int_{\mathbb{R}^2} f(\mintwodet | \boldsymbol{x}, s, \boldsymbol{c}_i) D(\boldsymbol{x}) f(s)d\boldsymbol{x}ds\).

Given our data, we obtain the following full likelihood \begin{equation}
\mathcal{L}_f( \boldsymbol{\phi}, \boldsymbol{\theta}, \boldsymbol{\eta}, \boldsymbol{\nu}, \kappa  |n, \boldsymbol{\Omega}, \boldsymbol{Y}, \boldsymbol{R}, \boldsymbol{C}) ,
\end{equation} where \(\boldsymbol{\phi}\) contains the parameters
associated with the specified density model for \(\boldsymbol{x}_i\),
\(\boldsymbol{\theta}= (\theta_U, \theta_B, \theta_Q)\) contains the
parameters associated with the Janoschek detection function,
\(\boldsymbol{\eta}= (\beta_r, \sigma_r)\) contains the parameters
related to received level, \(\boldsymbol{\nu}= (\mu_s, \sigma_s)\)
contains the parameters associated with the distribution of source
levels, and \(\kappa\) is the concentration parameter associated with
the bearings. Note that \(n, \boldsymbol{\Omega}, \boldsymbol{Y}\) and
\(\boldsymbol{R}\) now refer to the observed data and no longer to the
stochastic variables.

\newpage

\bibliography{references}